\documentclass[journal]{IEEEtran}

\usepackage{psfrag}
\usepackage{subfigure}
\usepackage{url}
\usepackage{stfloats}
\usepackage{cite}
\usepackage{amsmath}
\usepackage{amssymb}
\usepackage{amsfonts}
\usepackage{graphicx}
\usepackage{fancybox}
\usepackage{color}
\usepackage{algorithm}
\usepackage{algorithmic}
\usepackage{multirow}
\usepackage{upgreek}
\usepackage{booktabs}
\usepackage{threeparttable}
\usepackage{latexsym}
\usepackage{bm,array}
\usepackage{graphicx}
\usepackage{float}
\usepackage{mathtools}
\usepackage{commath}

\usepackage{stfloats}
\usepackage{cuted}
\usepackage{amsmath,lipsum}
\usepackage{cuted}

\IEEEoverridecommandlockouts
\normalsize
\begin{document}

\title{Distributed Reconfigurable Intelligent Surfaces for Energy Efficient Indoor Terahertz Wireless Communications}
\author{
Yiming~Huo,~\IEEEmembership{Senior Member,~IEEE},
Xiaodai~Dong,~\IEEEmembership{Senior Member,~IEEE}, and
Nuwan~Ferdinand,~\IEEEmembership{Member,~IEEE}
\thanks{Y. Huo, and X. Dong are with the Department of Electrical and Computer Engineering, University of Victoria, Victoria, BC V8P 5C2, Canada (ymhuo@uvic.ca, xdong@ece.uvic.ca). (Corresponding author: Xiaodai Dong)}
\thanks{N. Ferdinand is with Huawei Technologies, Ottawa, Canada (nuwan.ferdinand@huawei.com).}
}

\maketitle

\begin{abstract}
With the fifth-generation (5G) networks widely commercialized and fast deployed, the sixth-generation (6G) wireless communication is envisioned to provide competitive quality of service (QoS) in multiple aspects to global users. The critical and underlying research of the 6G is, firstly, highly dependent on the precise modeling and characterization of the wireless propagation when the spectrum is believed to expand to the terahertz (THz) domain. Moreover, future networks' power consumption and energy efficiency are critical factors to consider. In this research, based on a review of the fundamental mechanisms of reconfigurable intelligent surface (RIS) assisted wireless communications, we utilize the 3D ray-tracing method to analyze a realistic indoor THz propagation environment with the existence of human blockers. Furthermore, we propose a distributed RISs framework (DRF) to assist the indoor THz wireless communication to achieve overall energy efficiency. The numerical analysis of simulation results based on more than 2,900 indoor THz wireless communication sub-scenarios has demonstrated the significant efficacy of applying distributed RISs to overcome the mobile human blockage issue, improve the THz signal coverage, increase signal-to-noise ratios (SNRs), and QoS. With practical hardware design constraints investigated, we eventually envision how to utilize the existing integrated sensing and communication techniques to deploy and operate such a system in reality. Such a distributed RISs framework can also lay the foundation of efficient THz communications for Internet-of-Things (IoT) networks.       

\end{abstract}


\IEEEpeerreviewmaketitle
\newtheorem{mydef}{Definition}
\newtheorem{myLemma}{Lemma}
\newtheorem{theorem}{Theorem}
\newtheorem{Remark}{Remark}

\section{Introduction}
With the unprecedented speed of deployment of the fifth generation (5G) technologies by 2022, the new horizon of the next-generation wireless communications, or the sixth generation (6G), has gradually emerged and presented both challenges and opportunities \cite{Rappaport:6G}. The unexploited frequencies ranging from 100 GHz to 3 THz can be potentially used by the future 6G networks and systems. The candidate spectrum includes the THz region formally defined from 300 GHz to 3 THz. Unlike the THz radiation technology widely used for scientific research on medical imaging, security screening, and planetary exploration \cite{Huo:5GB}, THz wireless communications have been confronted with a series of challenges, although it brings more competitive benefits compared to the mmWave counterparts. The availability of a larger spectrum is a major advantage, which enables faster wireless data links. For example, 6G applications operating at THz bands may facilitate Tbps WLAN system (Tera-WiFi), Tbps Internet-of-Things (Tera-IoT), Tbps integrated access and backhaul (Tera-IAB) wireless networks, and ultra-broadband THz space communications (Tera-SpaceCom) \cite{Han:THz2019}. The smaller wavelength makes it feasible for the integration of ultra-massive multiple-in multiple-out (UM-MIMO) \cite{Han:UM-MIMO} into the THz communications system. 

However, the physical features of the THz wireless propagation are not thoroughly understood due to the lack of enough channel measurements at such high frequencies of interest, which stems from the unavailability and insufficient performance of the channel-sounding equipment and techniques. Authors in \cite{Priebe:THz} have presented one of the earliest indoor channel and propagation measurements for the line-of-sight (LOS) and non-line-of-sight (NLOS) paths at 300 GHz. In further research \cite{Khalid:THz}, path loss and phase delay were measured at a wide frequency range from 260 GHz to 400 GHz, with speeds of terabits per second demonstrated. Moreover, in a report summarizing the channel measurements campaign by researchers from NYU \cite{Rappaport:6G}, cross-polarization discrimination (XPD) and partition loss measurements at 28, 73, and 140 GHz for common indoor building materials. The scattering theory results were derived to help understand how frequency and surface roughness affect scattering behaviors across the mmWave and THz bands. Authors in \cite{Ju:ICC2019} provided reflection/scattering measurements at 140 GHz that can further develop the Directive Scattering (DS) models and theory. 

One critical yet largely underestimated factor is the THz channel modeling, requiring the precise and rigorous characterization of the electromagnetic (EM) wave propagation, which highly depends on accurately extracting the physical features of the wireless propagation environments, including building materials and atmospheric conditions. As early as 2005, \cite{Piesiewicz:2005} presented the absorption coefficient and refractive index measurements of typical building materials at THz bands. A further extension of the approach used in \cite{Piesiewicz:2005} is applied from the smooth building materials to the ones with rough surfaces characterized by the Rayleigh factor so that the scattering analysis can be obtained \cite{Piesiewicz:2007}. 
Furthermore, another recent research has developed a hybrid modeling approach of 3D ray-tracing simulations in a realistic office room at 300 and 350 GHz \cite{Sheikh:2020}.  

In addition, recent research advances in metamaterials, \cite{Pendry:Science} \cite{Schurig:Science}, have enabled the feasibility of applying the metasurface to improve wireless communications \cite{Nie:2018}. The concept of this new paradigm for wireless communications is based on the metasurface's capability to manipulate electromagnetic (EM) waves. It has evolved from the reconfigurable reflectarray/metasurface to the software-controlled metasurfaces, then the real-time software-controlled metasurfaces. \cite{Bjorson:Mag_2020}. Currently, it is well known as intelligent reflecting surface (IRS) \cite{Wu:IRS_2019} and reconfigurable intelligent surface (RIS). Authors in \cite{Nie:2018} have shown significant coverage improvement by using HyperSurfaces at 60 GHz. Research \cite{Pan:RIS} has enhanced the cell-edge user performance of multicell communication systems by employing an IRS at the cell boundary. Moreover, the signal interference can be largely mitigated using the RIS to reduce the random scatters in the wireless channels \cite{Nie:Nature}. 

Furthermore, a new hybrid wireless network comprising both active base stations (BSs) and passive IRSs has been studied in \cite{Lyu:2021}. Furthermore, the same team has investigated a multiuser system aided by multiple IRSs, and characterized its achievable spatial throughput averaged over channel fading and random IRS/UE locations \cite{Lyu:2020}. More recently, authors in \cite{Wu:WPNs} have proposed a novel dynamic IRS beamforming optimization framework. In particular, the system sum throughput of an IRS-aided wireless powered communication network (WPCN) can be maximized by jointly optimizing the IRS phase shifts and resource allocation for downlink wireless power transfer (DL WPT) and uplink wireless information transfer (UL WIT).

On the other hand, the efforts to put RISs from theory to practice have been witnessed. The proof of concept (PoC) design of several RIS testbeds working at sub-6 GHz and millimeter (mmWave) frequencies have been presented to validate several analytical path loss models \cite{Tang:TWC}, \cite{Tang:Refine}. Authors in \cite{Zhang:Nature} have demonstrated a 2-bit space-time-coding (STC) digital metasurface design for signal multiplexing in both the space and frequency domains. Two targeted users at different locations can receive the signal independently and simultaneously, while the undesired users at other locations are intentionally bypassed.   

In a typical THz wireless communication environment where multiple physical effects such as reflection and scattering can co-exist and lead to reduced received energy or interferences, the EM wave manipulating capability of the RIS and a controlled wireless communication channel is beneficiary. For example, the interference caused by the aforementioned physical effects can be mitigated while preserving the transmission energy in the desired spatial directions. As a result, the overall quality of service (Qos) and energy efficiency is improved. 

However, to the best of our knowledge, there is very little research work on analyzing and proposing how RIS(s) system should be designed and implemented in a cost-effective and energy-efficient way to assist the THz wireless communication in a realistic environment with practical design constraints and performance requirements. We believe a proper design and deployment of RIS(s) system is critical in improving the overall energy efficiency of the THz system and network, particularly when the state-of-the-art THz components and hardware are expensive with limited performances \cite{Hua:PA}.    

This paper proposes a solution framework based on the distributed RISs to enhance the THz wireless communications performance in a realistic indoor environment where blockage from both the static objects and mobile human persons exists. The indoor environment scenario is chosen for this research since it is more realistic for operating THz wireless communications with a comparatively more predictable atmospheric status (no rain, stable humidity). Also, it is more practical to deploy and maintain a RIS(s) system in the said environment. The most significant and practical concerns and challenges of enabling RIS-assisted indoor wireless communications are namely

\begin{itemize}
 \item \emph{\textbf{First}}, the RIS systematic energy efficiency, which is determined by both RIS system hardware design and also the RIS deployment and use strategies; 
 
 \item \emph{\textbf{Second}}, some critical indexes indicating the QoS such as SNR and latency need to be optimized; third, the QoS needs to be maintained when indoor persons behave as mobile blockage; 
 
 \item \emph{\textbf{Last but not least}}, RIS hardware aesthetic design should not be bulky and has a simple installation procedure. 
 
\end{itemize}


Our contributions are unfolded in several aspects:
\begin{enumerate}
\item \textcolor{black}{We first review some important radio propagation characteristics and analytical path loss models of RIS-assisted wireless communications. Then, we construct a realistic 3D indoor THz wireless communication application scenario where multiple user equipment and mobile human blockers exist. In particular, the materials and roughness factor of walls and furniture are considered and modeled.}

\item We develop a solution framework to overcome the THz signal dynamic blockage issues caused by a mobile human blocker(s) to provide significant SNRs improvement and thus maintain an ideal QoS. The key idea is to utilize multiple RISs specified with optimized dimensions and arranged in a distributed manner. More critically, an energy-efficient algorithm named ``ray searching and beam selecting'' is developed under this framework to operate the RISs collaboratively. As a result, the best system QoS can be achieved with minimized latency and power consumption.  

\item \textcolor{black}{Furthermore, to quantitatively analyze the performance of the proposed framework and conduct some possible adjustments if needed, we emulate the mobile human blockage scenarios by conducting spatial sampling. Subsequently, we successfully create abundant sub-scenarios to verify the framework and fine-tune its parameters if necessary. Eventually, we present data visualization to compare the performance gains under various combinations of application scenarios.}

\item \textcolor{black}{With practical hardware limitations and design constraints considered, such as the form factor and design complexity of the RIS system, we eventually provide an in-depth analysis of feasible design and deployment strategies of such a distributed RISs system.}

\end{enumerate}

The remainder of this article is organized as follows. Section II provides a brief review of radio propagation and absorption in-depth. Section III conducts an in-depth review of the physical features of the RIS and its analytical path loss models. Then, in Section IV, we present the distributed RISs framework (DRF) in a realistic indoor environment and the main idea of the ray searching and beam selecting algorithm. Furthermore, section V presents the quantitative analysis, data visualization, and comparison of the numerical results. Moreover, the design and deployment guidelines of a DRF system is proposed. Eventually, Section VI concludes this paper with future work discussed.   

\section{Physical Characteristics of RIS-Assisted Wireless Communications}

\subsection{Fresnel Reflection}
When an EM wave propagates in one medium and impinges on the surface of another medium with different electrical properties, how the EM wave will be partially reflected and transmitted depends on the surface of the medium \cite{Rappaport:2002}. Furthermore, whether the second medium is a perfect dielectric or a perfect conductor will determine how the EM wave is transmitted and reflected. For example, a perfect dielectric enables part of the energy to be transmitted to the second medium and part of the energy to be reflected back to the first medium without losing the energy in absorption. In contrast, a perfect conductor can reflect all the energy.

The Fresnel reflection coefficient is a function of the material properties, the frequency of the EM wave, wave polarization, and the angle of incidence. By using the Fresnel's equation, we can calculate the reflection coefficients of both transverse-electric (TE) and transverse-magnetic (TM) for the smooth surface when the EM wave propagates in the free space and impinges on a different material as follows \cite{Piesiewicz:2005}: 
\begin{equation}\label{eq:refl_te}
\gamma_{TE}=\frac{Z_{2}cos\theta_{i}-Z_{1}cos\theta_{t}}{Z_{2}cos\theta_{i}+Z_{1}cos\theta_{t}},
\end{equation}

and

\begin{equation}\label{eq:refl_tm}
\gamma_{TM}=\frac{Z_{2}cos\theta_{t}-Z_{1}cos\theta_{i}}{Z_{2}cos\theta_{t}+Z_{1}cos\theta_{i}},
\end{equation}
where $Z_{1}$ is the impedance of the first medium, and it can be the free-space wave impedance (377~$\Omega$) in some cases, while
$Z_{2}$ is the wave impedance of the second medium, while $\theta_{i}$ and $\theta_{t}$ represent the angle of incident and angle of refraction, respectively. The wave impedance of the reflecting material, $Z_{2}$, can be further calculated as \cite{Piesiewicz:2005}:
\begin{equation}\label{eq:impedance}
Z=\sqrt{\frac{\mu_{0}}{\varepsilon_{0}(n_{t}^2-(\alpha c/4\pi f)^2-j(2n_{t}\alpha c/4\pi f))}},
\end{equation}
where $\mu_{0}$ and $\varepsilon_{0}$ are the free-space permeability and permittivity, respectively. $c$ is the speed of light in the propagation medium. $\alpha$ is the absorption coefficient, and $n_{t}$ represents the refractive index which can be checked at frequencies of 100 GHz, 350 GHz, 500 GHz, and 750 GHz for several materials in\cite{Piesiewicz:2005} and \cite{Piesiewicz:2007}. In particular, $n_{t}$ is both frequency and material dependent, which can be checked from \cite{Piesiewicz:2007}, \cite{Jansen:2011}, and \cite{{Piesiewicz:2007_2}}, which will be used in the ray-tracing simulations.

\subsection{Absorption in THz Bands}
One significantly different feature of the THz radio propagation lies in its absorption rate for indoor THz communications, where the atmospheric condition (temperature, humidity, atmospheric pressure, etc.) is more controllable and stable than its outdoor counterpart. Therefore, thoroughly investigating and characterizing these two communication scenarios' impacting factors is necessary.   

\subsubsection{Atmospheric Gases}\hfill\\
Wherever the indoor or outdoor communication scenario applies, the atmospheric gases introduced attenuation is the basis which accumulates molecular absorption mainly from the oxygen, nitrogen, rare gases and water vapor. In terms of the ITU model \cite{ITU:Gas}, the specific gaseous attenuation is given by: 
\begin{equation}
\begin{aligned}
\label{eqn:gaseous}
\gamma&=\gamma_{o}+\gamma_{w}
\\&=0.1820 f_{G}\left(N_{\text {Oxygen }}^{\prime \prime}(f_{G})+N_{\text {Water Vapour}}^{\prime \prime}(f_{G})\right)\quad(\mathrm{dB}/\mathrm{km}),
\end{aligned}
\end{equation}
where $\gamma_{o}$ and $\gamma_{w}$ stand for the specific attenuation ($dB/km$) due to dry air (oxygen, pressure-induced nitrogen and non-resonant Debye attenuation) and water vapor, respectively. $f_{G}$ is the frequency in GHz, and $N_{\text {Oxygen }}^{\prime \prime}(f_{G})$ and $N_{\text {Water Vapour}}^{\prime \prime}(f_{G})$ are the imaginary parts of the frequency-dependent complex refractivities that can be further checked from \cite{ITU:Gas}, which are also functions about the dry air pressure, the water vapour pressure and the temperature.

\subsubsection{Specific Attenuation Models under Various Meteorological Conditions}\hfill\\
In this sub-section, we present the three major models under rain, fog, and snow, respectively. First, in terms of the specification attenuation model for the rain proposed by the ITU, the rain rate indexed by $R$ (mm/h) will determine how much attenuation is obtained per kilometer:
\begin{equation}
\begin{aligned}
\label{eqn:gammarain}
\gamma_\text{R}=k R^{\alpha} ,
\end{aligned}
\end{equation}
where the frequency-dependent parameters, $k$ and $\alpha$ are determined by equations that are developed from curve-fitting to power-law coefficients derived from scattering calculations \cite{ITU:Rain}.

Furthermore, fog (or cloud) is a visible aerosol composed of tiny water droplets or ice crystals (less than 0.01 cm) suspended in the air at or near the Earth's surface. It can cause significant impacts on outdoor THz communications. The specific attenuation (dB/km) within a fog or cloud is formulated as \cite{ITU:Fog}:
\begin{equation}
\begin{aligned}
\label{eqn:gammafog}
&\gamma_{\text{c}}(f, T)=K_{l}(f, T) M\quad(dB/km), 
\end{aligned}
\end{equation}
where $K_{l}$ is the specific attenuation coefficient (($dB/km$)/($\mathrm{g}/\mathrm{m}^{3})$), which can be calculated as equation (\ref{eqn:fogkl}). $M$ is the density of liquid water in the cloud or fog ($\mathrm{g}/\mathrm{m}^{3}$). Fog attenuation can be very significant at frequencies around 100 GHz or above. For medium fog, the density of liquid water in the fog is usually about 0.05 $\mathrm{g}/ \mathrm{m}^{3}$ (visibility is about 300 $m$), and dense fog is 0.5 $\mathrm{g}/ \mathrm{m}^{3}$ (visibility is about 50 $m$).  
\begin{equation}
\begin{aligned}
\label{eqn:fogkl}
K_{\text{l}}(f, T)=\frac{0.819 f_{G}}{\varepsilon^{\prime \prime}\left(1+\eta^{2}\right)} \quad(\mathrm{dB} /\mathrm{km})/\left(\mathrm{g}/ \mathrm{m}^{3}\right),
\end{aligned}
\end{equation}
where $\eta=\frac{2+\varepsilon^{\prime}}{\varepsilon^{\prime \prime}} $ , and the complex permittivity of water, $\varepsilon^{\prime}$ and $\varepsilon^{\prime \prime}$ can be further calculated in terms of \cite{ITU:Fog}.
Finally, the specific attenuation caused by snow is based on the model in\cite{Oguchi:Electromagnetic}.  In terms of the specific liquid content, snow can be divided into dry snow and wet snow \cite{Koch:Snow}, and this article only considers the specific attenuation model of the dry snow as follows:
\begin{equation}
\begin{aligned}
\label{eqn:gammasnow}
\gamma_{\text{s}}=0.00349 \frac{R_{\text{s}}^{1.6}}{\lambda^{4}}+0.00224 \frac{R_\text{s}}{\lambda} \quad(dB/km) ,
\end{aligned}
\end{equation}
where $R_{\text{s}}$ is the snowfall speed in millimeters per hour while $\lambda$  is the wavelength in centimeters. 

\subsection{Fundamental Mechanism of the RIS}
Metasurfaces' superior capability of tailoring EM waves makes integrating RIS(s) into indoor environments to assist THz wireless communications hold great potential for better performance. First, the unique physical features of the RIS need to be thoroughly investigated, reviewed, and analyzed. 

\begin{figure}
\centering
\includegraphics[scale = 0.52]{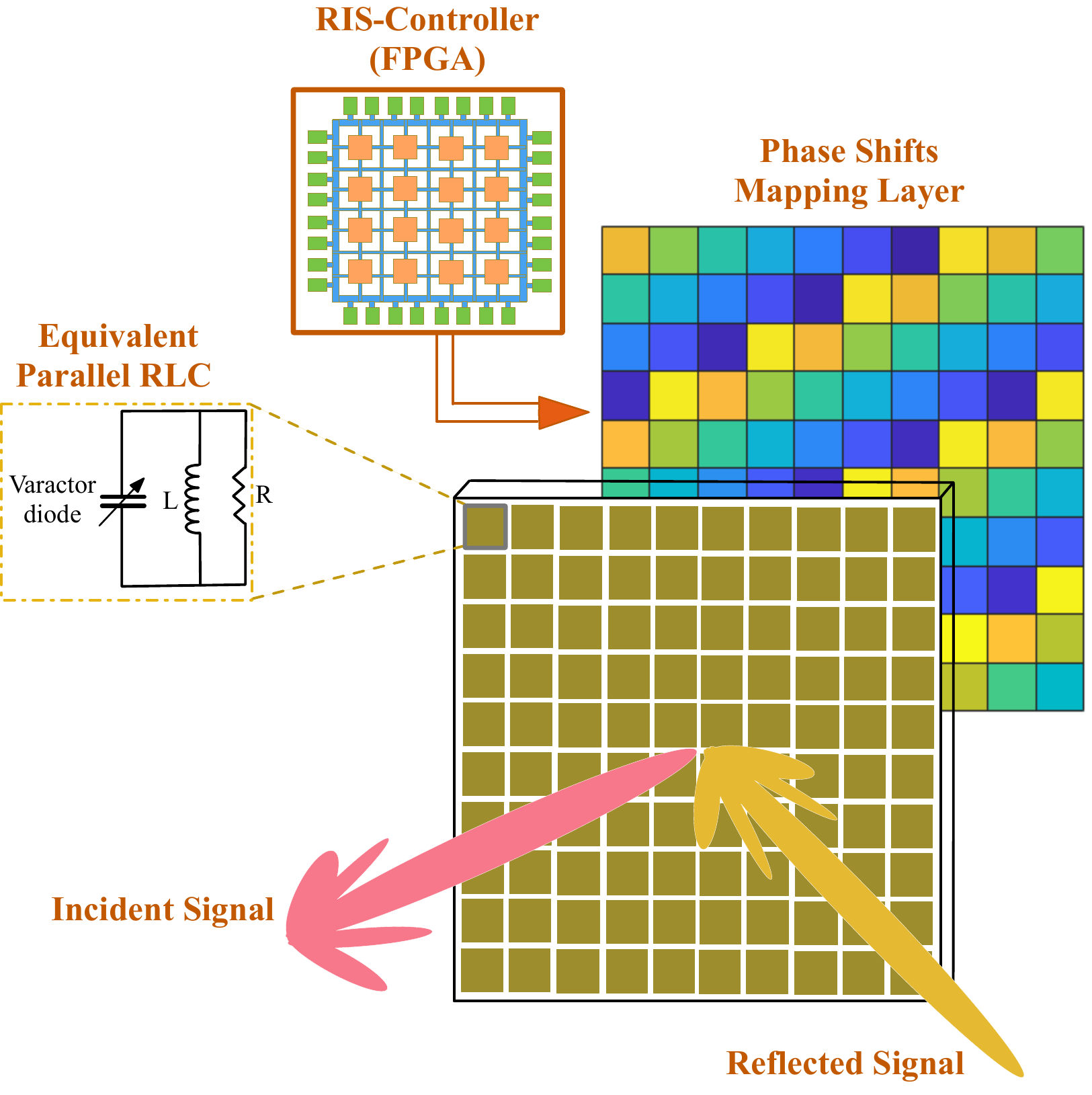}
\caption{The block diagram of a typical reconfigurable intelligent surface (RIS) system example.}\label{fig:RIS_BLOCK}
\end{figure}

Unlike the reflectarrays that are designed to form a planar phase surface in front of the aperture with fixed radiation patterns \cite{Pozar:Reflectarray}, a typical RIS consists of a large number of sub-wavelength-sized RIS elements (also called unit cells, meta cells, meta atoms, etc.) acting as diffuse scatterers, as illustrated by an example in Fig.~\ref{fig:RIS_BLOCK}. From a microscopic scale, each RIS element, i.e., the scatter, is connected to a varactor diode (with electrically tunable capacitance). Subsequently, the state, e.g., amplitude and phase of each RIS element, can be altered. More specifically, each RIS unit cell can be modeled with an equivalent RLC tank where one or more varactor diodes provide variable capacitance. The varactor diode is popularly used in voltage-controlled oscillator (VCO) designs \cite{Huo:VCO} for frequency tuning. As a result, the load impedance, $Z_\text{L}$, of the RLC tank representing the RIS unit cell is voltage-controlled; therefore, the reflection coefficient of the RIS unit cell is also voltage-controlled and can be written as \cite{Pozar:Book}

\begin{equation}
\varGamma_\text{RIS-cell}=\frac{Z_\text{L}-Z_\text{0}}{Z_\text{L}+Z_\text{0}},
\end{equation}
where $Z_\text{0}$ is the characteristic impedance of the free space. In addition, the reflection phase tuning is derived as
\begin{equation}
\varphi(\varGamma_\text{RIS-cell})=arctan\left(\frac{Im(\varGamma_\text{RIS-cell})}{Re(\varGamma_\text{RIS-cell})}\right).
\end{equation}

Furthermore, all RIS cells' varactor diodes are controlled by corresponding discrete bias voltages of a RIS controller. A popular implementation manner is to use the field programmable gate array (FPGA), digital-to-analog converter (DAC) module, or shift registers so that the real-time programming of the RIS can be realized \cite{Tang:CC}, \cite{Bjornson:Prototyping}. As illustrated in Fig. \ref{fig:RIS_BLOCK}, the phase shifts are realized by the RIS-controller and correspond to each RIS cell with a phase shift within 0 to 2$\pi$ (dark blue to light yellow).   

\begin{figure}
\centering
\includegraphics[scale = 0.55]{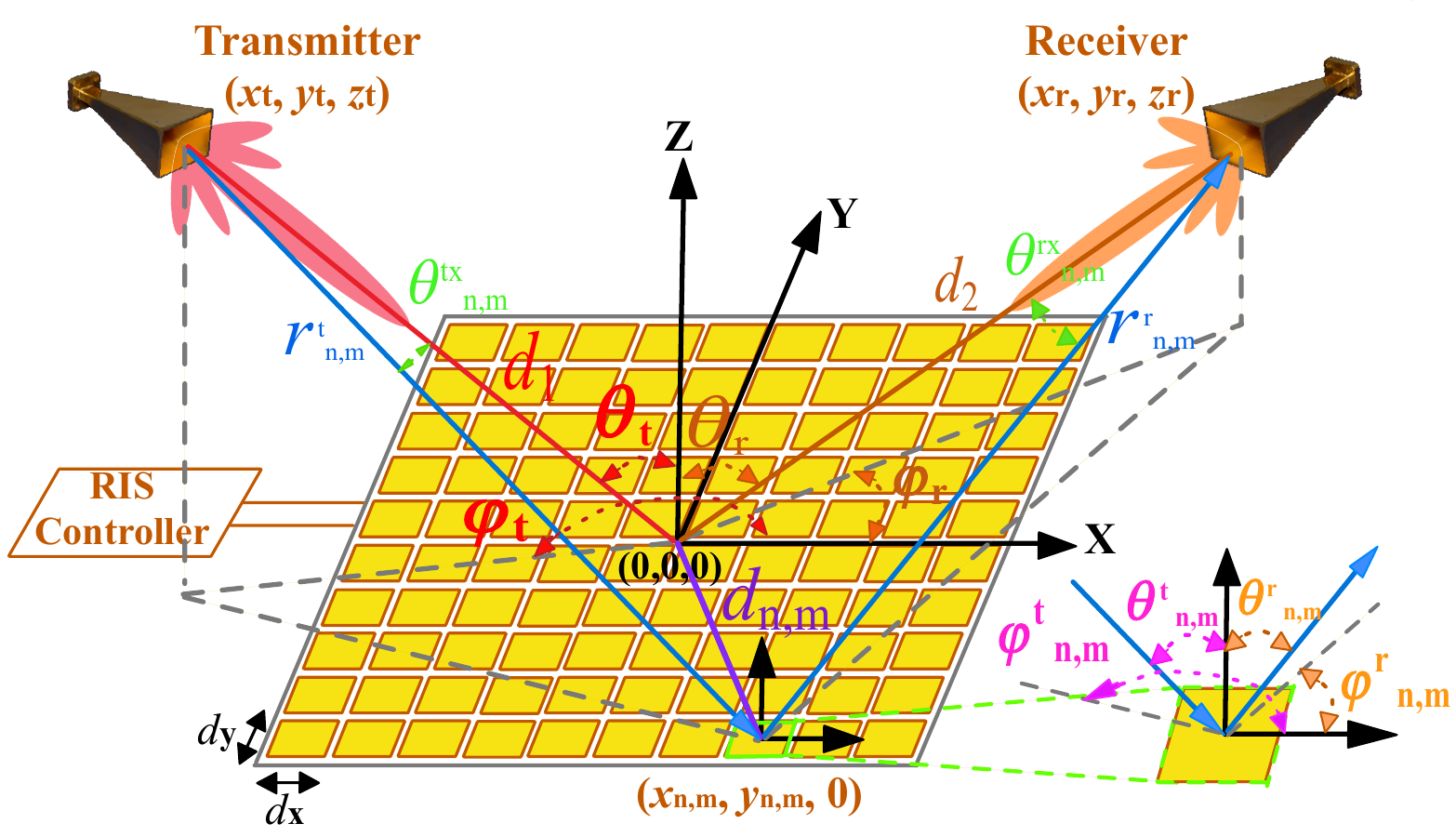}
\caption{RIS-assisted wireless communication without LOS path.}\label{fig:RIS_PL}
\end{figure}

It is also worth mentioning that each RIS element should be substantially smaller than the wavelength to scatter signals more uniformly. Therefore the RIS can form equally strong radiation at all observation angles \cite{Bjornson:Myth}, \cite{Ozdogan:RIS}. Also, in a metasurface, the dimensions and spacing between the RIS elementary elements are typically between $\lambda/10$ and $\lambda/5$ \cite{Tapio:RIS}, \cite{Glybovski:RIS}. However, in a realistic RIS prototyping design, the dimension of a RIS unit cell is subject to a handful of practical hardware design trade-offs and optimization, therefore the actual adopted dimensions are diverse. In most existing RIS designs, the two-dimensional spacings of RIS cells, $d_{x}$ and $d_{y}$ are both $\leq \lambda/2$. For example, authors have specified a fifth of the wavelength in the metasurface design \cite{Tsilipakos:RIS} while researchers from Southeast University set ($d_{x}$, $d_{y}$) to (0.35$\lambda$, 0.35$\lambda$) at 10.5 GHz, (0.17$\lambda$, 0.17$\lambda$) at 4.25 GHz, (0.126$\lambda$, 0.252$\lambda$) at 27 GHz, (0.418$\lambda$, 0.418$\lambda$) at 33 GHz respectively in their PoC demos \cite{Tang:TWC}, \cite{Tang:Refine}. As observed, implementing sufficiently small RIS elements is desired but can increase the I/O interface complexity and cost of the RIS controller.

\subsection{Path Loss Models of RIS-Assisted Wireless Communication}
The RIS-assisted wireless communications path loss models depend on the Fraunhofer distance, which is defined as the boundary of the far field and the near field of the antenna array with the following expression
\begin{equation}\label{eq:Fraunhofer}
L=\frac{2D^{2}}{\lambda},
\end{equation}
where $L$ and $D$ denote the distance between the transmitter and antenna array's center and the largest dimension of the antenna array, respectively. Moreover, assuming there are $N$ rows and $M$ columns of RIS unit cells evenly distributed in the RIS system, with (0, 0, 0) as the origin of the Cartesian coordinate system, as illustrated in Fig.~\ref{fig:RIS_PL}, the reproduced figure of the conceptual demonstration in \cite{Tang:TWC}. Therefore, (\ref{eq:Fraunhofer}) can be rewritten as
\begin{equation}\label{eq:Fraunhofer2}
L=\frac{2MNd_{x}d_{y}}{\lambda}.
\end{equation}
Next, the general free-space path loss model of the RIS-assisted wireless communication is derived as (\ref{eq:PL_Gen}) from \cite{Tang:Refine}, shown at the bottom of this page. In (\ref{eq:PL_Gen}), $P_{t}$ is the power emitted from the transmitter, $P_{r}$ stands for the received power at the receiver, $G_{t}$, $G_{r}$ are the antenna gains (in linear scale) at transmitter and receiver, respectively. $r_{n,m}^{t}$ and $r_{n,m}^{r}$ represent the travelled distances between the transmitter/receiver and the RIS unit cell on the $n^{th}$ row and $m^{th}$ column, noted as $U_{n,m}$. $\varGamma_{n,m}$ denotes the reflection coefficient of the RIS unit cell. 

Furthermore, $F_{n,m}^{combine}$ is the joint normalized power radiation pattern defined as (\ref{eq:fcombine}) from \cite{Tang:Refine}, shown at the bottom of this page. In (\ref{eq:fcombine}), $F^{tx}(\theta, \varphi)$, $F^{rx}(\theta, \varphi)$, and $F(\theta, \varphi)$ represent the normalized radiation patterns of the TX antenna, the RX antenna, and the individual RIS unit cell, respectively. In this paper, we adopt the same general model as \cite{Tang:Refine} for the normalized power radiation pattern (NPRP) of a transmit antenna as follows


\begin{figure*}[!b]
\begin{equation}
\begin{aligned} \label{eq:PL_Gen}
PL_\text{general}=\frac{P_{t}}{P_{r}}=\frac{16\pi^{2}}{G_{t}G_{r}(d_{x}d_{y})^{2}\abs{\sum\limits_{m=1}^{M} \sum\limits_{n=1}^{N} \frac{\sqrt{F_{n,m}^{combine}}\varGamma_{n,m}}{r_{n,m}^{t}r_{n,m}^{r}}e^{\frac{-j2\pi(r_{n,m}^{t}+r_{n,m}^{r})}{\lambda}}}^{2}},
\end{aligned}
\end{equation}
\end{figure*}

\begin{figure*}[!b]
\begin{equation}
\begin{aligned} \label{eq:fcombine}
F_{n,m}^{combine}=F^{tx}(\theta_{n,m}^{tx},\varphi_{n,m}^{tx})F(\theta_{n,m}^{t},\varphi_{n,m}^{t})F(\theta_{n,m}^{r},\varphi_{n,m}^{r})F^{rx}(\theta_{n,m}^{rx},\varphi_{n,m}^{rx}),
\end{aligned}
\end{equation}
\end{figure*}

\begin{equation}\label{eq:nprp}
F^{tx}\left( {\theta ,\varphi } \right) = \left\{
\begin{array}{rcl}
{\left(\cos\theta\right)}^\alpha & & {\theta  \in \left[ {0,\frac{\pi }{2}} \right], \varphi  \in \left[ {0,2\pi } \right]}\\
0 & & {\theta  \in \left( {\frac{\pi }{2},\pi } \right], \varphi  \in \left[ {0,2\pi } \right]}.
\end{array} \right.
\end{equation}  

Furthermore, by taking into account the TX/RX antenna gains, $G_{t}$ and $G_{r}$, we have the following NPRP expressions \cite{Tang:Refine}
\begin{equation}\label{eq:nprp_tx}
F^{tx}\left( {\theta ,\varphi } \right) = \left\{
\begin{array}{rcl}
{\left(\cos\theta\right)}^{(\frac{G_t}{2} - 1)} & & {\theta  \in \left[ {0,\frac{\pi }{2}} \right], \varphi  \in \left[ {0,2\pi } \right]}\\
0 & & {\theta  \in \left( {\frac{\pi }{2},\pi } \right], \varphi  \in \left[ {0,2\pi } \right]}.
\end{array} \right.
\end{equation}

\begin{equation}\label{eq:nprp_rx}
F^{rx}\left( {\theta ,\varphi } \right) = \left\{
\begin{array}{rcl}
{\left(\cos\theta\right)}^{(\frac{G_r}{2} - 1)} & & {\theta  \in \left[ {0,\frac{\pi }{2}} \right], \varphi  \in \left[ {0,2\pi } \right]}\\
0 & & {\theta  \in \left( {\frac{\pi }{2},\pi } \right], \varphi  \in \left[ {0,2\pi } \right]}.
\end{array} \right.
\end{equation}
Additionally, we use the same NPRP expression for each individual RIS unit cell as in \cite{Tang:Refine}
\begin{equation}\label{eq:nprp_cell}
F\left( {\theta ,\varphi } \right) = \left\{
\begin{array}{rcl}
{\cos\theta} & & {\theta  \in \left[ {0,\frac{\pi }{2}} \right], \varphi  \in \left[ {0,2\pi } \right]}\\
0 & & {\theta  \in \left( {\frac{\pi }{2},\pi } \right], \varphi  \in \left[ {0,2\pi } \right]}.
\end{array} \right.
\end{equation}

\begin{figure*} [!t]
\begin{equation}
\begin{aligned} \label{eq:fcombine2}
&F_{n,m}^{combine}\mathop  = {\left( {\cos \theta _{n,m}^{tx}} \right)^{^{(\frac{{{G_t}}}{2} - 1)}}}\left( {\cos \theta _{n,m}^t} \right)\left( {\cos \theta _{n,m}^r} \right){\left( {\cos \theta _{n,m}^{rx}} \right)^{^{(\frac{{{G_r}}}{2} - 1)}}} \\
&= {\left( {\frac{{{{\left( {{d_1}} \right)}^2} + {{\left( {r_{n,m}^t} \right)}^2} - {{\left( {{d_{n,m}}} \right)}^2}}}{{2{d_1}r_{n,m}^t}}} \right)^{(\frac{{{G_t}}}{2} - 1)}}\left( {\frac{{{z_t}}}{{r_{n,m}^t}}} \right)\left( {\frac{{{z_r}}}{{r_{n,m}^r}}} \right){\left( {\frac{{{{\left( {{d_2}} \right)}^2} + {{\left( {r_{n,m}^r} \right)}^2} - {{\left( {{d_{n,m}}} \right)}^2}}}{{2{d_2}r_{n,m}^r}}} \right)^{(\frac{{{G_r}}}{2} - 1)}}.
\end{aligned}
\end{equation}
\end{figure*}


It is worth noting that, the aforementioned NPRP equations have been verified by the experimental measurement results and should be sufficiently general \cite{Tang:Refine}. Subsequently, by substituting the aforementioned NRRP equations and some further geometric derivation into (\ref{eq:fcombine}) and assuming the directions of peak radiation of both TX and RX antennas point towards the center of the RIS, we can obtain~(\ref{eq:fcombine2}) from \cite{Tang:Refine}, shown at the top of the next page. In (\ref{eq:fcombine2}), $d_{1}$ and $d_{2}$ stand for the distances between the transmitter/receiver and the RIS's geometric center, i.e. (0, 0, 0), $d_{nm}$ indicates the distance between $U_{n,m}$ and (0, 0, 0), which are illustrated in Fig. \ref{fig:RIS_PL}. Assume that the amplitude of the reflection coefficient of all RIS unit cells, $\varGamma_{n,m}$, is identical and equal to $A$. The free-space path loss of a RIS-assisted link in the far-field beamforming scenario is as follows

\begin{equation}\label{eq:PL_FarBeam}
\begin{aligned}
PL_{farfield}^{beam}=\frac{{16{\pi ^2}{{({d_1}{d_2})}^2}}}{{{G_t}{G_r}{{\left( {MN{d_x}{d_y}} \right)}^2}F({\theta _t},{\varphi _t})F({\theta _r},{\varphi _r}){A^2}}} \\
= \frac{{16{\pi ^2}{{({d_1}{d_2})}^2}}}{{{G_t}{G_r}{{\left( {MN{d_x}{d_y}} \right)}^2}{\cos {\theta _t}}{\cos {\theta _r}}{A^2}}}.
\end{aligned}
\end{equation}
Moreover, still assuming that the amplitude of the reflection coefficient of all the unit cells is the same, i.e., $\left| {\varGamma _{n,m}} \right|=A$, the free-space path loss model for RIS-assisted communications in the near-field beamforming scenario can be expressed as
\begin{equation}\label{eq:PL_NearBeam}
\begin{aligned}
PL_{nearfield}^{beam}=\frac{P_t}{P_r}=\frac{{16{\pi ^2}}}{{{G_t}{G_r}{\left({d_x}{d_y}\right)}^2{A^2}{{\left| {\sum\limits_{m = 1}^{M} {\sum\limits_{n = 1}^{N} { \frac{\sqrt {{F_{n,m}^{combine}}}}{{r_{n,m}^tr_{n,m}^r}}} } } \right|}^2}}},
\end{aligned}
\end{equation} 
and this formula applied to the situation that the TX and RX are both or only one of them is in the near-field region of the RIS. On top of the same situation, when the RISs are electrically large, we have the specific free-space path loss model for RIS-assisted near-field broadcasting \cite{Tang:TWC} as follows

\begin{equation}\label{eq:broadcast}
PL_{nearfield}^{broadcast} \approx \frac{{16{\pi ^2}{{({d_1} + {d_2})}^2}}}{{{G_t}{G_r}{\lambda ^2}{A^2}}}.
\end{equation}

Additionally, \cite{Tang:Refine} has introduced the free-space path loss model for the single RIS unit cell as
\begin{equation}\label{eq:single}
\begin{aligned}
PL_{U_{n,m}}&=\frac{P_t}{P_{n,m}^r}\\
&= \frac{{16{\pi ^2}}}{{{G_t}{G_r}{{\left( {{d_x}{d_y}} \right)}^2}{{\left| {\frac{{\sqrt {F_{n,m}^{combine}}\ {\varGamma_{n,m}}}}{{r_{n,m}^tr_{n,m}^r}}{e^{\frac{{ - j2\pi (r_{n,m}^t + r_{n,m}^r)}}{\lambda }}}} \right|}^2}}}\\
&= \frac{{16{\pi ^2}{{\left( {r_{n,m}^tr_{n,m}^r} \right)}^2}}}{{{G_t}{G_r}{{\left( {{d_x}{d_y}} \right)}^2}F_{n,m}^{combine}{{\left| {{\varGamma_{n,m}}} \right|}^2}}}.
\end{aligned}
\end{equation}

Equations (\ref{eq:PL_Gen}), and (\ref{eq:PL_FarBeam})--(\ref{eq:single}) derived from \cite{Tang:TWC} and \cite{Tang:Refine} demonstrate the free-space path loss models under several situations without considering the atmospheric attenuation which is needed to include when calculating the realistic path loss.  

\begin{figure*} 
\begin{minipage}{.5\linewidth}
\centering
\subfigure[]{\includegraphics[scale=.62]{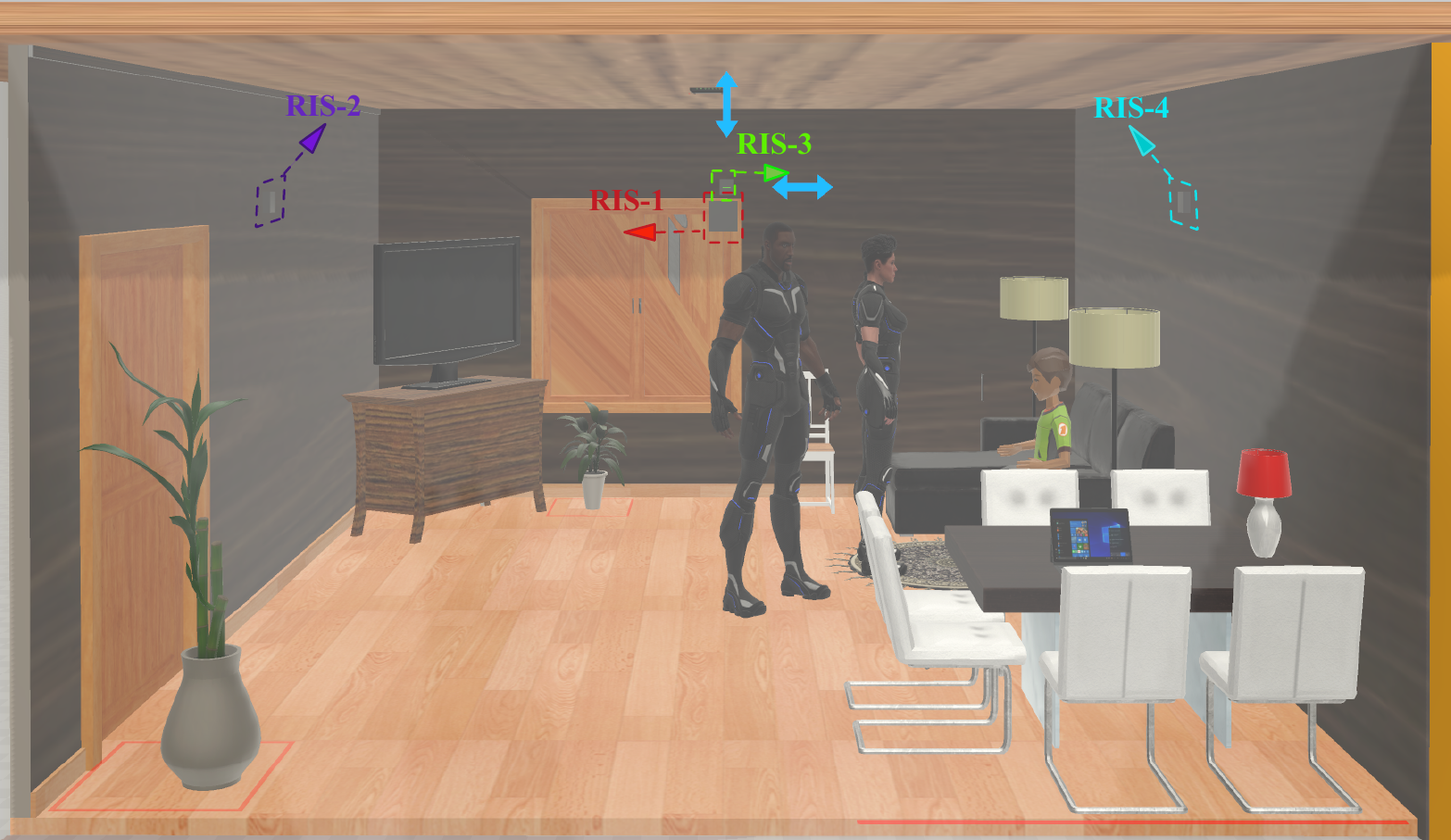}}
\end{minipage}
\begin{minipage}{.5\linewidth}
\centering
\subfigure[]{\includegraphics[scale=.46]{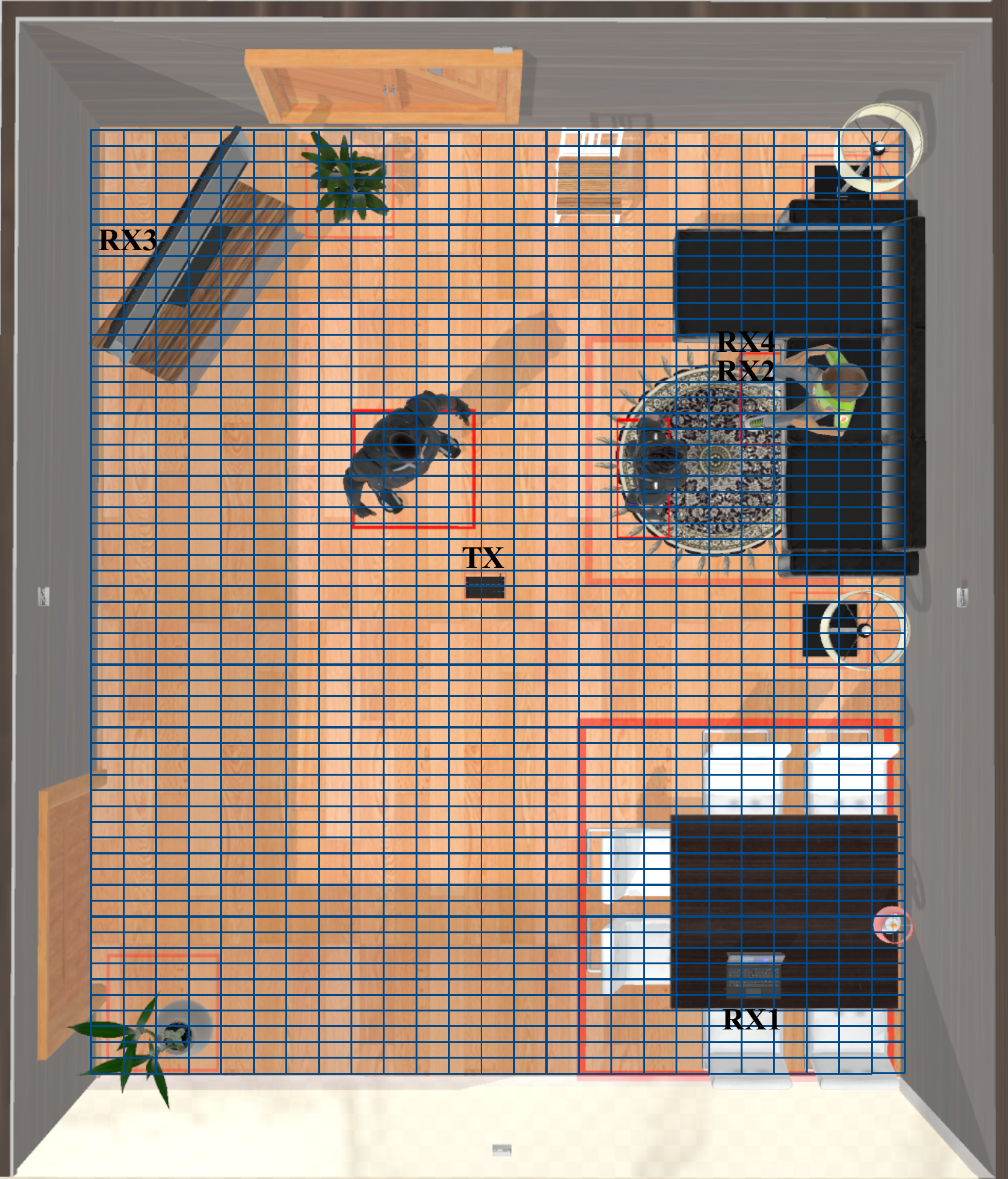}}
\end{minipage}\par\medskip
\caption{Illustration of (a) the distributed RISs framework (DRF) enabled indoor THz Wireless communication (transparent view through the front wall) and (b) the top-down transparent view of the DRF-enabled indoor THz Wireless communication with the mask layer of spatial sampling.} \label{fig:Indoor3D}
\end{figure*}

\section{Distributed RISs for THz Indoor Wireless Communications}
\subsection{Motivation and Design Target}

It is worth noting there are some interesting features and conclusions observed and obtained from the path loss models of the RIS-enabled wireless communication. 

\begin{itemize}
 \item First, assume that in a far-filed beamforming situation where the total distance of the route from the TX to RISs plus RISs to the RX, $d_\text{total}$ (=$d_\text{1}+d_\text{2}$) is fixed, the path loss peaks when $d_\text{1} = d_\text{2}$ and decreases when either $d_\text{1}$ or $d_\text{2}$ gets small. This is to say, the path loss will be minimized when placing the RISs close to the source or destination, which is also discussed in \cite{Bjornson:Power}, \cite{Bjornson:RIS2}.
 \item Moreover, the RISs' total hardware area is inversely proportional to the path loss with a roll-off factor at 20~dB/octave. However, the hardware area is practically limited by the circuitry complexity, hardware cost, power consumption, and the product design aesthetic, which needs to be compatible with the indoor environment. For example, a large room may need the RIS system sufficiently large to overcome the long transmission distance. Still, the hardware area cannot be too large due to the aforementioned design considerations.
 \end{itemize}
 
Consequently, distributing multiple RISs of suitable dimensions and realistic hardware design for the indoor environment can play a critical role in enabling good THz wireless communication performance with practical design requirements satisfied. Presumably, the proposed distributed RISs system needs to deliver several specifications and features as follows:

 \begin{enumerate}
\item Being able to integrate fully and be compatible with the indoor environment without significantly impacting the decoration and building architecture. This at least implies the RIS hardware area and product volume need to be well maintained, e.g., comparable to a credit card or smartphone (0.15 $\times$ 0.1 m$^{2}$). 

\item Enabling high network quality of service, particularly when mobile human blockage exists. For example, a large attenuation of 45.6 dB has been observed in the experiment at 310 GHz, when the transmission is obstructed by a human hand \cite{Khalid:THz2}. Indoor human activities can block the direct LOS propagation, offering the shortest transmission distance and the highest signal-to-noise ratio (SNR). Moreover, the signal strengths of multi-path components after the second or higher-order bounce (reflection) and traveling are much weaker than direct propagation or the channel component(s) experiencing only one bounce in terms of the THz channel measurement and modeling. The RIS-reconstructed channel propagation needs to enable the considerably good performance of SNR.     

\item Furthermore, defining the good SNR performance obtained through using the distributed RIS framework (DRF) is a prerequisite. In this paper, we propose that when using the DRF to enable a RIS-reconstructed THz radio propagation in an indoor environment with no LoS communication, the resulting SNR needs to be close to or outperform the SNR of the direct propagation from TX to RX.

\item The proposed DRF can fast detect the channel and determine which RIS can enable the receiver's best reception and establish the said communication link accordingly. The proposed DRF can effectively handle the mobile human blockage to enhance the QoS. More critically, it can do so with higher energy efficiency, which is critical to extending the RIS's long operation duration regardless RIS is power-supplied by battery, AC wall power, wireless charging, solar power, etc.     
\end{enumerate}


\begin{algorithm}[t!]
{\footnotesize\caption{\footnotesize Ray Searching and Beam Selecting for Distributed RISs Framework}
\label{alg:algorithm_DRF}
\hspace{\algorithmicindent} \textbf{Input: } Indoor room 3D design; TX coordinate ($x_\text{TX}, y_\text{TX}, z_\text{TX}$); RXs coordinates: $\mathbf{(x_\text{RX}, y_\text{RX}, z_\text{RX})}$; Single RIS entity parameters: $N, M, d_\text{x}, d_\text{y}$, $\varGamma_\text{n,m}$; number of RISs: $N_\text{RISs}$; building material refractive index: $n_{t}$, carrier frequency: $f_{c}$; atmospheric condition: $T, P, W$; TX/RXs antenna(s) gains: $G_\text{TX}$, $G_\text{RX}$;\\
\hspace{\algorithmicindent} \textbf{  Output: } $\mathbf{SNR_\text{Gain-LOS}}$, $\mathbf{SNR_\text{Gain-NLOS}}$, $\mathbf{J_\text{RIS}}$;
	\begin{algorithmic}[1]
		\STATE Generate all $N_\text{bloc}$ human blockage positions based on the indoor 3D design file using the proposed spatial sampling method, the $i^\text{th}$ human position is $(x_\text{bloc}(i), y_\text{bloc}(i))$; 
		\STATE Generate all $N_\text{RISs}$ RIS surfaces' coordinates $\mathbf{(x_\text{RISs}, y_\text{RISs}, z_\text{RISs}) }$; 
		\FOR {$i = 1:  N_\text{bloc}$ }
		\STATE Indoor 3D ray tracing with blockage at $(x_\text{bloc}(i), y_\text{bloc}(i))$;  \newline
		Obtain the status of LOS communication ray, $N_\text{LOS}$ (either 0 or 1);\newline
		Obtain the number of NLOS communication rays, $N_\text{NLOS}$;  \newline 
		Obtain coordinates of all reflection points on walls/ceiling/floor; \newline
		Calculate the LOS traveled distances, $r_\text{LOS}$ and path loss $PL_\text{LOS}$; \newline
		Obtain all available traveled distances $\mathbf{r_\text{NLOS}}$ and calculate the corresponding path losses $\mathbf{PL_\text{NLOS}}$ 
		\FOR{$j = 1:  N_\text{RISs}$ }
		\STATE Indoor 3D ray tracing between RX/TX and $\text{RISs}(j)$; 
        \IF{LOS communication exists} 
        \STATE Obtain distances, $r_\text{TX-RISs(j)}$ and $r_\text{RISs(j)-RX}$ and store them;
        \IF{$r_\text{TX-RISs(j)}$ or $r_\text{RISs(j)-RX}>L$ in (\ref{eq:Fraunhofer})} 
        \STATE Calculate the incident/reflected angles of $\text{RISs}(j)$; \newline
        Far field beamforming is applied, (\ref{eq:PL_FarBeam}) and (\ref{eqn:gaseous}) are used to calculate $PL_{INDR-FAR}^{beam}$ and store the value;
        \ELSIF{$r_\text{TX-RISs(j)}$ and $r_\text{RISs(j)-RX}<L$} 
        \STATE Near field beamforming is applied, (\ref{eq:PL_NearBeam}) and (\ref{eqn:gaseous}) are used to calculate $PL_{INDR-NEAR}^{beam}$ and store the value;
        \ENDIF \newline
        Store the value of $j$ into $\mathbf{J}$ (a $N_\text{bloc}\times N_\text{RISs}$ matrix); \newline
        Compare the RIS-reconstructed link's path loss, $PL_\text{RIS}$ ($PL_{INDR-FAR}^{beam}$ or $PL_{INDR-NEAR}^{beam}$) with $P_\text{LOS}$ and path losses $\mathbf{PL_\text{NLOS}}(i,:)$;\newline
        $SNR_\text{Gain-LOS}$ = $PL_\text{LOS}$ - $PL_\text{RIS}$;\newline
        $SNR_\text{Gain-NLOS} = min(\mathbf{PL_\text{NLOS}}(i,:)) - PL_\text{RIS}$;\newline
        Store $SNR_\text{Gain-LOS}$, $SNR_\text{Gain-NLOS}$ into $\mathbf{SNR_\text{Gain-LOS}}(i,j)$, $\mathbf{SNR_\text{Gain-NLOS}}(i,j)$;
        \ENDIF
		\ENDFOR
		\STATE 
		Obtain the corresponding $j$ value for max($\mathbf{SNR_\text{Gain-LOS}}(i,:)$), max($\mathbf{SNR_\text{Gain-NLOS}}(i,:)$). Store $j$ value into $\mathbf{J_\text{RIS}}(i,1)$ (a $N_\text{bloc}\times 1$ vector);
		\ENDFOR
\end{algorithmic}}
\end{algorithm}

\subsection{Ray Searching and Beam Selecting}
To design and verify the DRF system that can effectively work in the most realistic indoor environment, Microsoft 3D Builder was firstly employed to design and develop a standard sitting room with a dimension of about $6 \times 5 \times 2.5~\text{m}^{3}$. As illustrated in Fig. \ref{fig:Indoor3D}, there are indoor walls, floor, ceiling, furniture, plants, home appliances, electronic devices, three persons, and a normal atmospheric condition that all become part of the THz radio propagation environment. In particular, a femtocell (e.g., access point (AP)) operating at THz bands is placed at the geometric center of the ceiling, and one RIS is attached to each of the four walls with some height. The femtocell operates as the central processing unit for signal processing and RISs control, e.g., phase shifts, through wired cables or wireless control links.      
To emulate the mobile human blockage and investigate how it affects the THz channel and radio propagation, the spatial sampling is conducted by placing on the indoor floor a mask layer consisting of a large number of small rectangular grids with a $0.1~\text{m} \times 0.2~\text{m}$ dimension. As shown in Fig.~\ref{fig:Indoor3D} (a), the grided floor plan resembles a chessboard. Next, we place the human blocker in the initial position by setting his/her coordinate to the geometric center of the corresponding small `chessboard' grid. Furthermore, we conduct the 3D ray-tracing simulation, calculate and record all critical channel parameters such as the number of available rays (LOS and NLOS), each ray's path loss, angle of arrival (AOA), angle of departure (AOD), reflection points' coordinates, channel delays, etc. 

Afterward, we move the human blocker from one small grid to another and repeat the same ray-tracing simulation and data recording. Eventually, all available human blockage positions (except those small grids occupied by furniture, appliances, etc) should be simulated and stored with a high spatial resolution. In the designed sitting room environment, there are totally 727 grids simulated to guarantee a high-accuracy channel characterization. Furthermore, the RISs' channel parameters at each human blockage position are recorded and analyzed to develop the DRF algorithm.

As illustrated in Algorithm \ref{alg:algorithm_DRF}, at the initial step, all $N_\text{bloc}$ human blockage positions are generated from the indoor 3D design file and all $N_\text{RISs}$ RIS surfaces' coordinates are specified. For the next,  the 3D ray tracing function is repeatedly operated for the mobile human blockage position  $(x_\text{bloc}(i), y_\text{bloc}(i))$ when $i$ increases from 1 to $N_\text{bloc}$ (727 in our design case). In the $i^\text{th}$ (human blockage position) situation, assume that there is no RIS existing, critical geometric parameters of the indoor environment are extracted and important channel state parameters are calculated, such as the LOS communication availability and its practical path loss $PL_\text{LOS}$, the number of available NLOS communication rays and the corresponding practical path loss(es) $\mathbf{PL_\text{NLOS}}(i,:)$.  

Furthermore, after obtaining and storing the parameters and data from the above indoor environment hypothetically without RIS assistance, the 3D ray-tracing function is executed again for the indoor environment with RISs to obtain the geometric relationship and calculate the channel parameters between each RIS and the transmitter/receiver. Subsequently, the LOS communication availability between each RIS and TX/RX is alternatively examined. The algorithm will calculate the communication distance of each RIS that is LoS-available to determine whether the near-field or far-field beamforming applies. and then use the corresponding equations, (\ref{eq:PL_FarBeam})/(\ref{eq:PL_NearBeam}) and (\ref{eqn:gaseous}), to calculate each activated RIS's actual  $PL_\text{RIS}$ and store it and its numbering. 

Moreover, the algorithm further compares each $PL_\text{RIS}$ with $PL_\text{LOS}$ and $\mathbf{PL_\text{NLOS}}(i,:)$ in the $i^\text{th}$ situation. Eventually, two types of SNR gains are calculated, namely $SNR_\text{Gain-LOS}$, the SNR gain from the LOS communication comparison and $SNR_\text{Gain-NLOS}$, the one from being compared to the best (smallest) NLOS ray's path loss, $min(\mathbf{PL_\text{NLOS}}(i,:))$. Finally, both SNR gains are stored into matrices $\mathbf{SNR_\text{Gain-LOS}}$ and $\mathbf{SNR_\text{Gain-NLOS}}$, and the numbering of the RIS leading to the best SNR gains is stored in a vector $\mathbf{J_\text{RIS}}$.

\begin{table}
\centering
\footnotesize
\caption{Simulation Parameters.}\label{Table:Simulation}
\begin{tabular}{|c|c|}
\hline
\textbf{Parameters} & \textbf{Settings}  \\
\hline
Environment & indoor sitting room \\
\hline
Temperature, pressure, humidity & 293.15 K, 101.325  kPa, 43\%  \\
\hline
Room dimension & $6 \times 5 \times 2.5~\text{m}^{3}$ \\
\hline
Surface materials & plasterboard \cite{Jansen:2011} \\
\hline
Carrier frequency $f_\text{c}$ & 300 GHz, 700 GHz \\
\hline
Antenna(s) gain $G_\text{TX}$, $G_\text{RX}$ & 20~dB, 10~dB \\
\hline
TX position & ceiling center, 2.45~m high \\
\hline
RX-1 position & laptop, 0.91~m high  \\
\hline
RX-2 position & user equipment 1, 0.95~m high  \\
\hline
RX-3 position & TV set-top-box, 1~m high  \\
\hline
RX-4 position & user equipment 2, 1.2~m high  \\
\hline
Human blockage positions, $N_\text{bloc}$ & 727 \\
\hline
Total number of RISs, $N_\text{RISs}$ & 4 \\
\hline
$d_{x}$, $d_{y}$ of each RIS & $0.35\lambda$, $0.35\lambda$  \\
\hline
$M$, $N$ of each RIS & adjustable variable \\
\hline
\end{tabular}
\end{table}

\begin{table*}[h]
\scriptsize
\caption{Statistical analysis of application scenarios without and with RISs operating at 300 GHz.} \label{tab:APP300G}
\newcommand{\tabincell}[2]{\begin{tabular}{@{}#1@{}}#2\end{tabular}}
 \centering
 \begin{threeparttable}
 \begin{tabular}{|c|c|c|c|c|c|c|c|c|c|c|}\hline
        \tabincell{c}{\textbf{}} & \multicolumn{8}{c|}{\textbf{Channel parameters extracted from simulations (both $\mu$ and $\sigma$ are in dB)}} & \tabincell{c}{~} & \tabincell{c}{~}     \\  \hline
        \tabincell{c}{\textbf{Application} \\ \textbf{Scenarios}}  & \multicolumn{2}{c|}{\tabincell{c}{\textbf{Without RISs}}}   & \multicolumn{2}{c|}{\tabincell{c}{\textbf{With RISs} \\ \textbf{Option-1}}} & \multicolumn{2}{c|}{\tabincell{c}{\textbf{With RISs} \\ \textbf{Option-2}}} &
        \multicolumn{2}{c|}{\tabincell{c}{\textbf{With RISs} \\ \textbf{Option-3}}} &
        \tabincell{c}{\textbf{Strategy}, \\ $\overline{\Delta\textbf{SNR}_\textbf{all}}$, \\ $\overline{\Delta\textbf{SNR}_\textbf{NLOS}}$ \\ \textbf{(dB)} } &
        \tabincell{c}{\textbf{Prob.} \\ \textbf{of former} \\ \textbf{NLOS} \\ \textbf{cases} \\ \textbf{improved}} \\  \hline
        \tabincell{c}{~} & \tabincell{c}{$\text{Occ.}_\text{LOS}$ \\($\mu_\text{PL}$, $\sigma_\text{PL}$)} & \tabincell{c}{$\text{Occ.}_\text{NLOS}^\text{min}$ \\($\mu_\text{PL}$, $\sigma_\text{PL}$)} & 
        \tabincell{c}{(RIS\#, Occ.) \\($\mu_\text{PL}$, $\sigma_\text{PL}$)} &
        \tabincell{c}{(RIS\#, Occ.) \\($\mu_\text{PL}$, $\sigma_\text{PL}$)} &
        \tabincell{c}{$\text{Occ.}_\text{LOS}$ \\($\mu_\text{PL}$, $\sigma_\text{PL}$)} &
        \tabincell{c}{(RIS\#, Occ.) \\($\mu_\text{PL}$, $\sigma_\text{PL}$)} &
        \tabincell{c}{(RIS\#, Occ.) \\($\mu_\text{PL}$, $\sigma_\text{PL}$)} &
        \tabincell{c}{$\text{Occ.}_\text{LOS}$ \\($\mu_\text{PL}$, $\sigma_\text{PL}$)} &
        \tabincell{c}{~} & \tabincell{c}{~} \\  \hline
        \tabincell{c}{RX1} & \tabincell{c}{98.9\% \\ (91.51, 0)} & \tabincell{c}{1.1\% \\ (102, 1.41)} &
        \tabincell{c}{(1, 99.72\%) \\ (89.39, 0)} & \tabincell{c}{(3, 0.28\%) \\ (90.17, 0)} &
        \tabincell{c}{98.9\% \\ (91.51, 0)} & \tabincell{c}{(1, 1.1\%) \\ (89.39, 0)} &
        \tabincell{c}{(1, 99.72\%) \\ (89.39, 0)} & \tabincell{c}{0.28\% \\ (91.51, 0)}
         &\tabincell{c}{Option-1, \\2.24,\\12.61} & \tabincell{c}{8/8,\\
         100\%}    \\ 
        \hline
        \tabincell{c}{RX2} & \tabincell{c}{N.A.} & \tabincell{c}{100\% \\ (102.04, 0.28)} &
        \tabincell{c}{(1, 99.86\%) \\ (85.82, 0)} & \tabincell{c}{(4, 0.14\%) \\ (104.22, 0)} &
        \tabincell{c}{N.A.} & \tabincell{c}{N.A.} &
        \tabincell{c}{(1, 99.86\%) \\ (85.82, 0)} & \tabincell{c}{*0.14\% \\ (102.04, 0)}
         &\tabincell{c}{Option-3, \\16.2,\\16.2} & \tabincell{c}{726/727,\\
         99.86\%}    \\ 
        \hline 
        \tabincell{c}{RX3} & \tabincell{c}{96.42\% \\ (91.22, 0)} & \tabincell{c}{3.58\% \\ (103.95,3.76)} &
        \tabincell{c}{(3, 98.35\%) \\ (89.04, 0)} & \tabincell{c}{(4, 1.65\%) \\ (102.19, 0)} &
        \tabincell{c}{96.42\% \\ (91.22, 0)} & \tabincell{c}{(1, 3.58\%) \\ (89.39, 0)} &
        \tabincell{c}{(3, 98.35\%) \\ (89.04, 0)} & \tabincell{c}{1.65\% \\ (91.22, 0)}
         &\tabincell{c}{Option-3, \\2.6,\\14.91} & \tabincell{c}{26/26,\\
         100\%}    \\ 
        \hline   
        \tabincell{c}{RX4} & \tabincell{c}{99.72\% \\ (90.1, 0)} & \tabincell{c}{0.28\% \\ (102.32, 0)} &
        \tabincell{c}{(1, 99.86\%) \\ (86.34, 0)} & \tabincell{c}{(3, 0.14\%) \\ (89.59, 0)} &
        \tabincell{c}{99.72\% \\ (90.1, 0)} & \tabincell{c}{(1, 0.14\%) \\ (86.34, 0) \\ + \\ (3, 0.14\%) \\ (89.59, 0) } & \tabincell{c}{N.A.} & \tabincell{c}{N.A.} &\tabincell{c}{Option-1, \\3.79,\\14.36} & \tabincell{c}{1/1,\\
         100\%}    \\ 
        \hline   
    \end{tabular}
    \begin{tablenotes}
        \footnotesize
        \item[*] There is no LOS but NLOS occurrence only.
      \end{tablenotes}
    \end{threeparttable}
\end{table*}

\begin{figure}
\begin{minipage}{.5\linewidth}
\centering
\subfigure[]{\includegraphics[scale=.52]{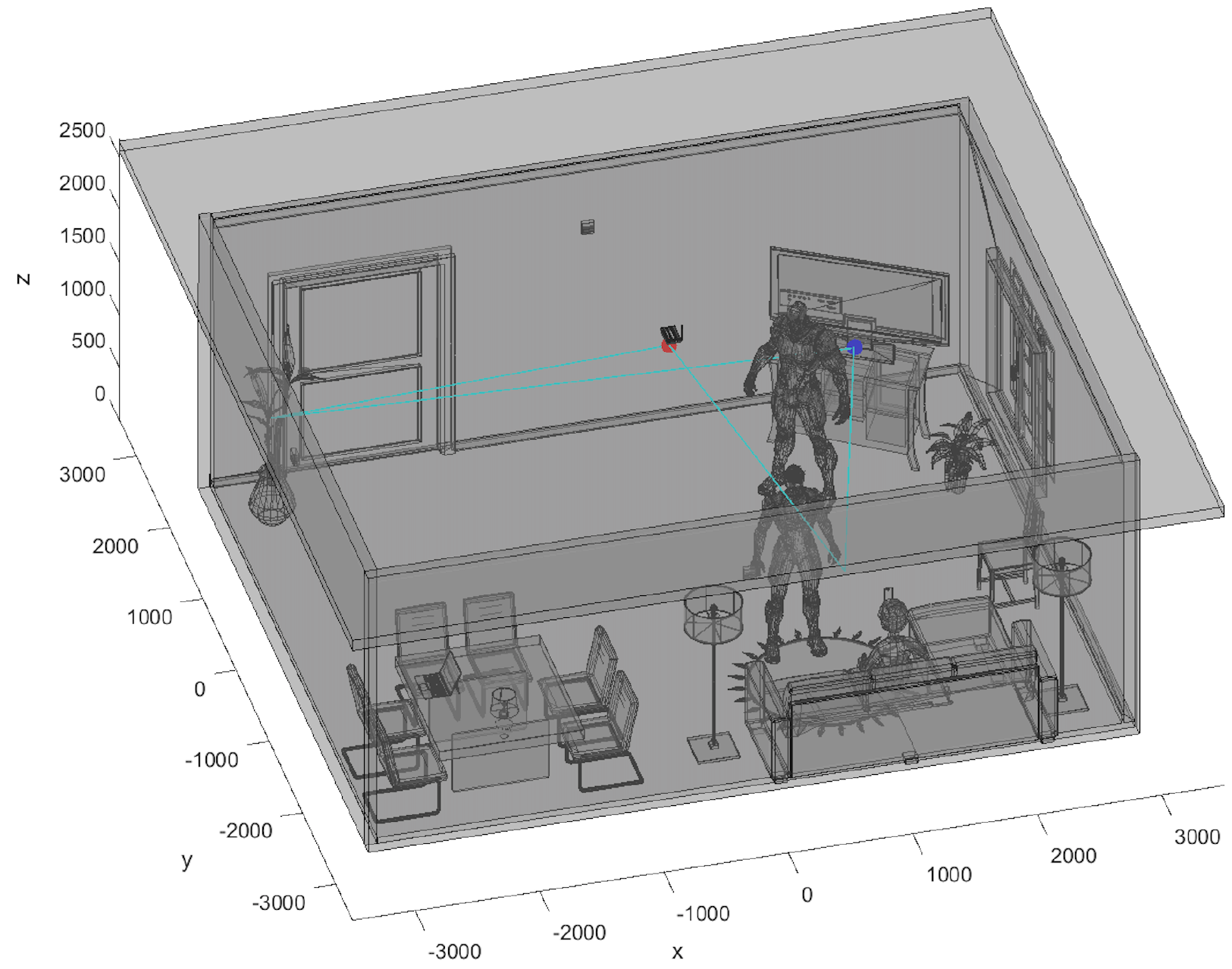}}
\end{minipage} \\
\begin{minipage}{.5\linewidth}
\centering
\subfigure[]{\includegraphics[scale=.52]{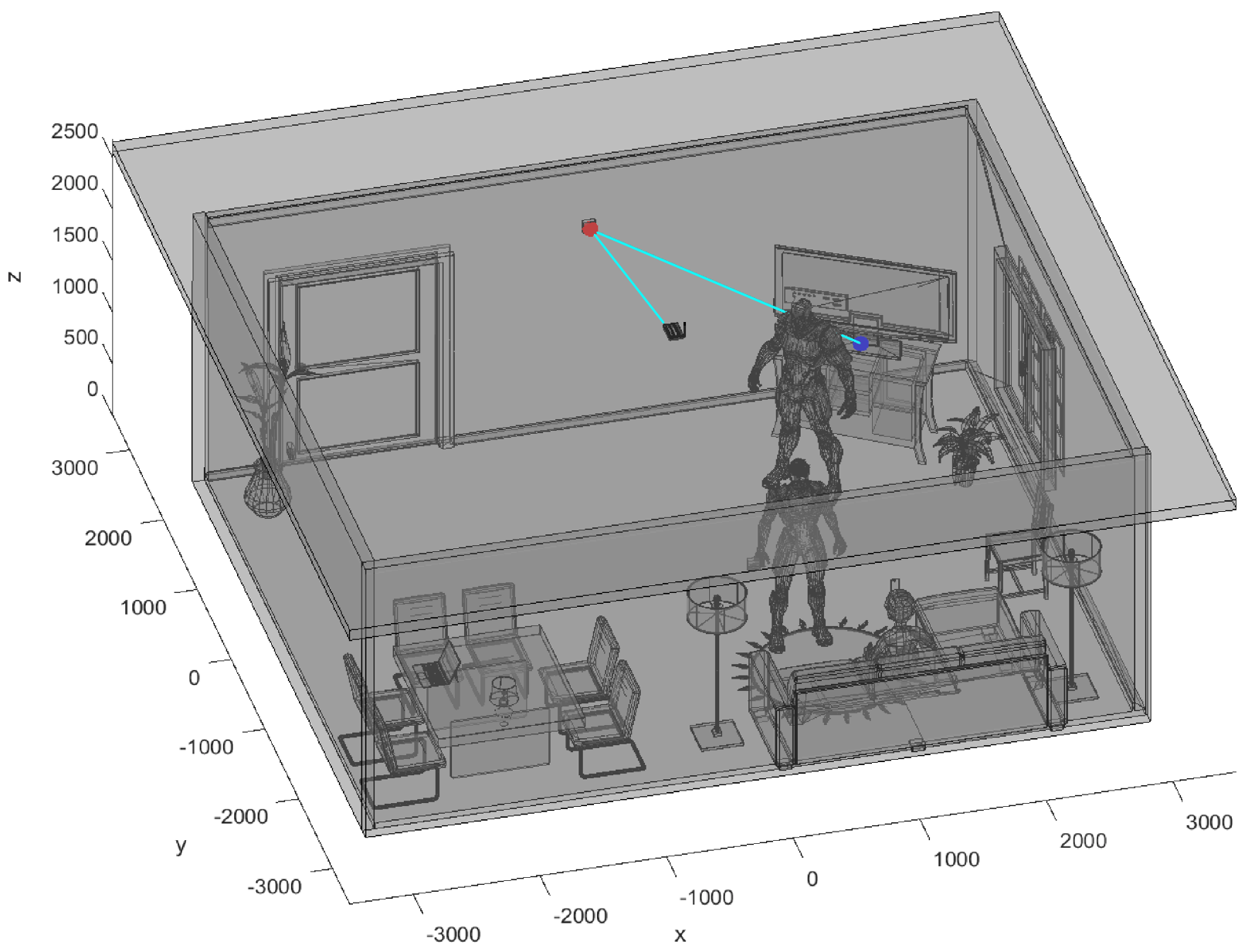}}
\end{minipage} \\
\begin{minipage}{.5\linewidth}
\centering
\subfigure[]{\includegraphics[scale=.52]{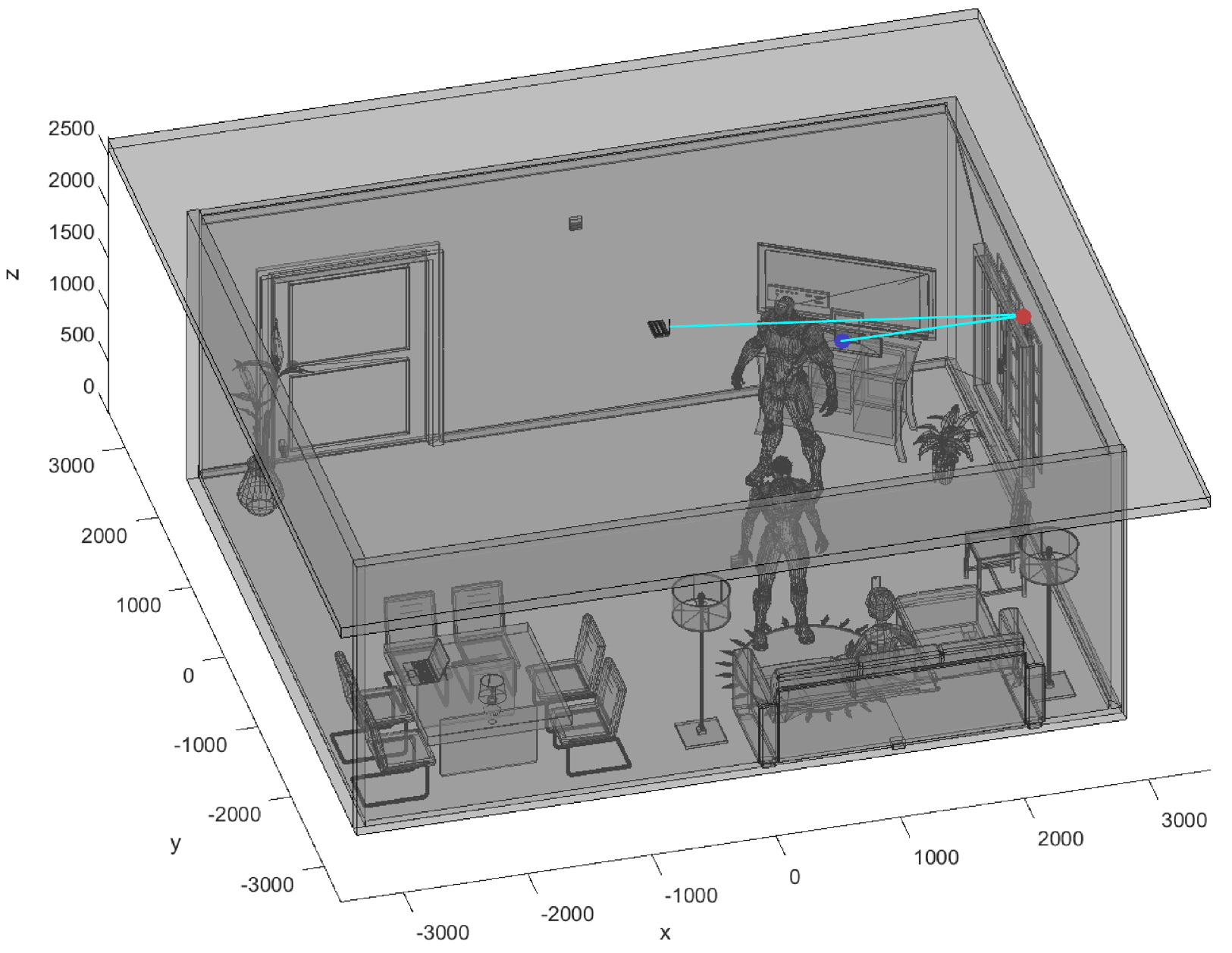}}
\end{minipage}\par\medskip
\caption{ Illustration of ray-tracing simulation results based on the human blockage position and RX-1 (a) without any RIS activated. DRF Algorithm 1 examines to find all available RISs and obtains the parameters (b) with RIS-3 assisted, and (c) with RIS-4 activated.}\label{fig:TV}
\end{figure}

\begin{figure}
\centering
\subfigure[]{\label{}\includegraphics[scale=.36]{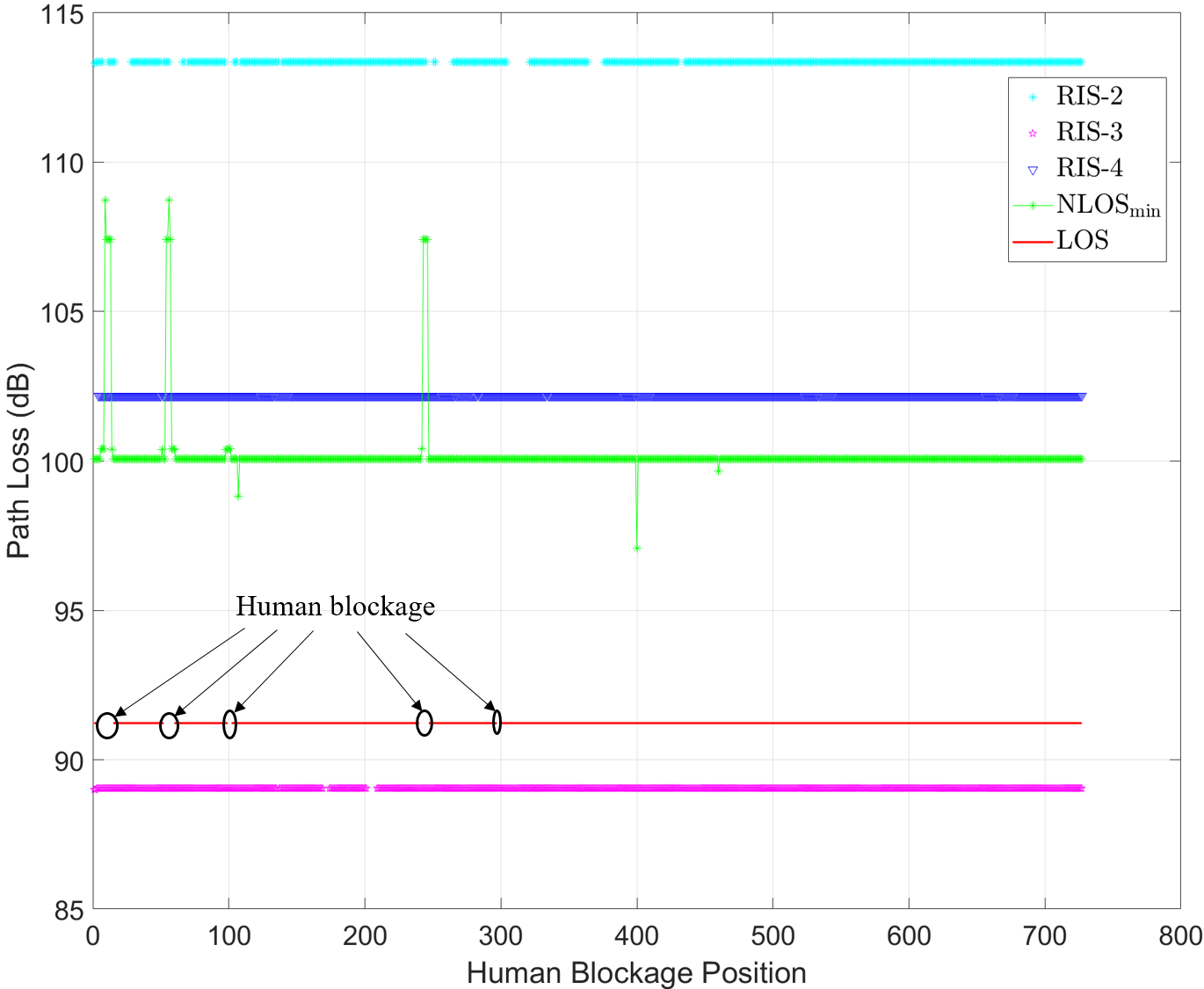}}

\begin{minipage}{.5\linewidth}
\centering
\subfigure[]{\includegraphics[scale=.3]{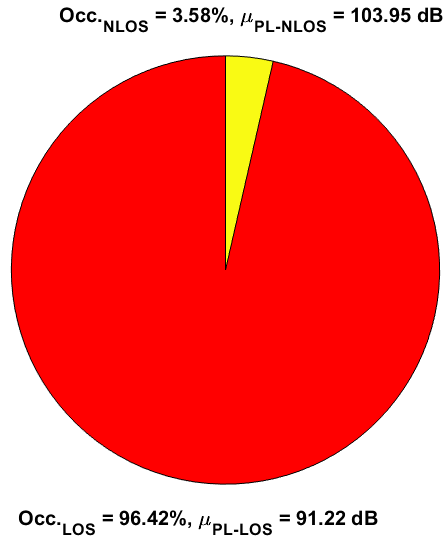}}
\end{minipage}%
\begin{minipage}{.5\linewidth}
\centering
\subfigure[]{\includegraphics[scale=.3]{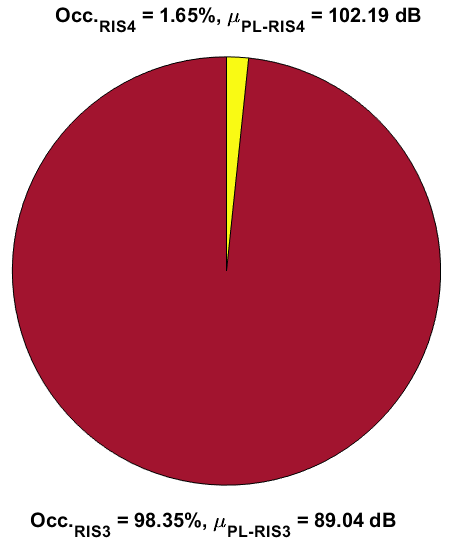}}
\end{minipage}\par\medskip

\begin{minipage}{.5\linewidth}
\centering
\subfigure[]{\includegraphics[scale=.3]{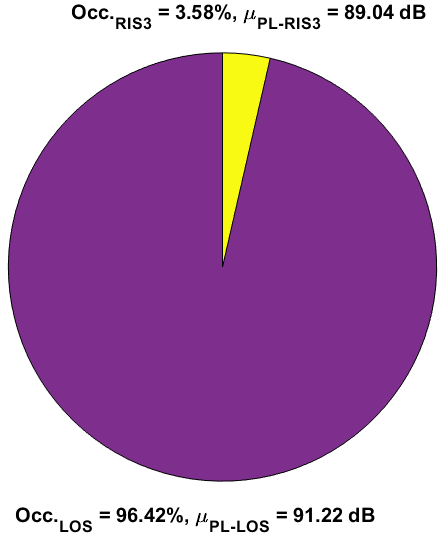}}
\end{minipage}%
\begin{minipage}{.5\linewidth}
\centering
\subfigure[]{\includegraphics[scale=.3]{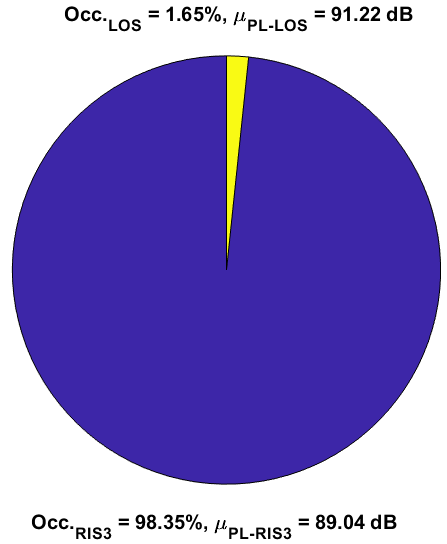}}
\end{minipage}\par\medskip
\caption{Statistical data visualization of ray-tracing simulation results based on all human blockage positions with 300 GHz carrier frequency and RX-3 as the receiver, including (a) path losses with all walls/floor/ceiling and RISs examined, and statistics of (b) the original situation without RISs, (c) replying on RIS-3 and RIS-4 only, (d) using RIS-3 only when blockage happens, (e) mainly using RIS-3 and switching to LOS when RIS-3 is unblocked.}\label{fig:HBP_RX-3}
\end{figure}

\begin{figure}

\centering
\subfigure[]{\label{}\includegraphics[scale=.38]{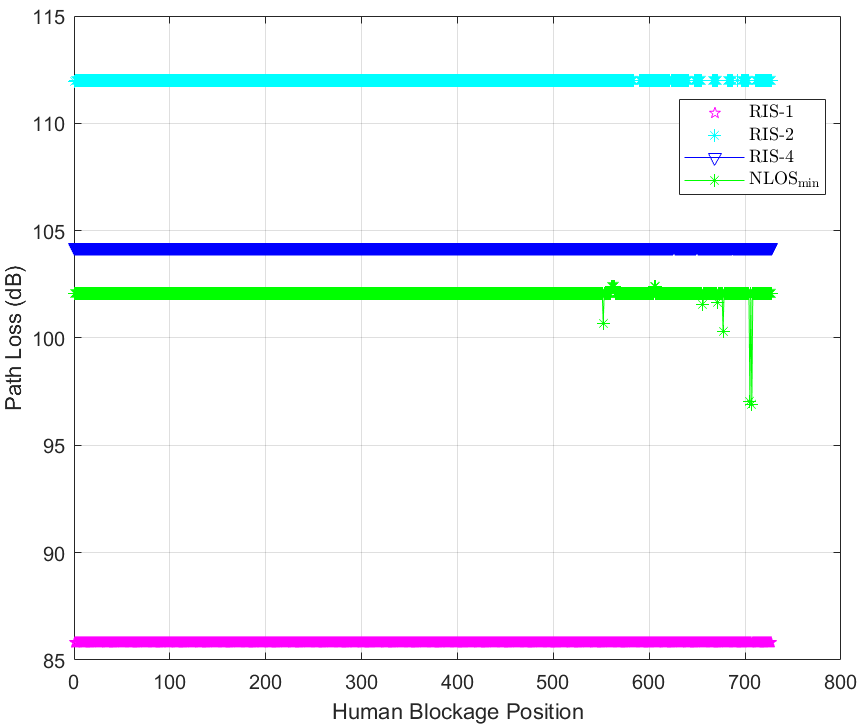}}

\begin{minipage}{.5\linewidth}
\centering
\subfigure[]{\includegraphics[scale=.3]{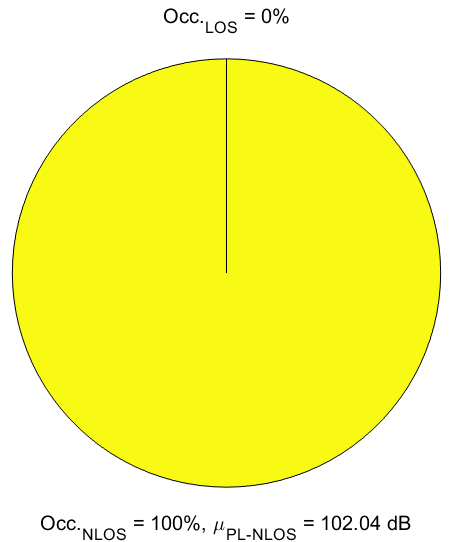}}
\end{minipage}%
\begin{minipage}{.5\linewidth}
\centering
\subfigure[]{\includegraphics[scale=.3]{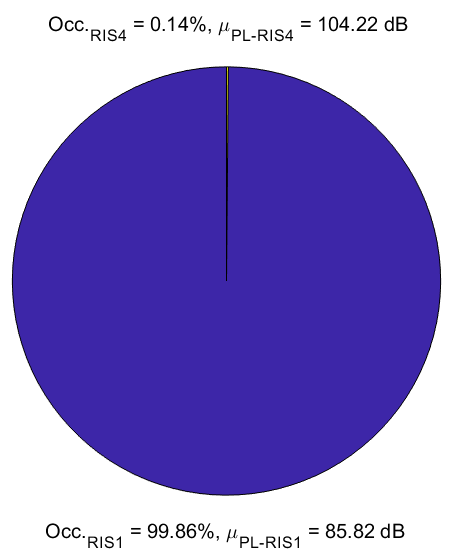}}
\end{minipage}\par\medskip

\begin{minipage}{.5\linewidth}
\centering
\subfigure[]{\includegraphics[scale=.3]{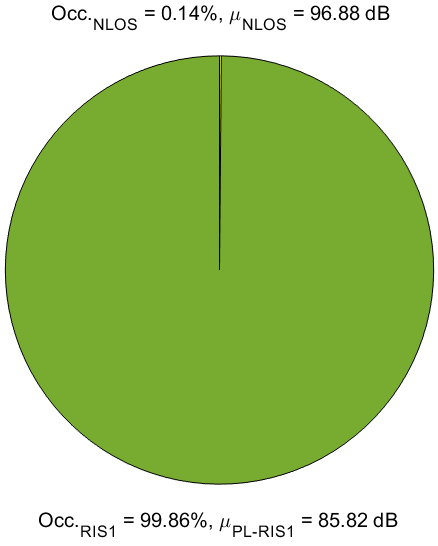}}
\end{minipage}\par\medskip
\caption{Statistical data visualization of ray-tracing simulation results based on all human blockage positions operating at 300 GHz and RX-2 as the receiver, including (a) path losses with all RISs examined, and statistics of (b) the original situation without RISs, (c) only relying on RIS-1 and RIS-4, (d) using RIS-1 only when blockage happens.}\label{fig:HBP_RX-2}
\end{figure}

\begin{figure*}
\centering
\includegraphics[scale = 0.52]{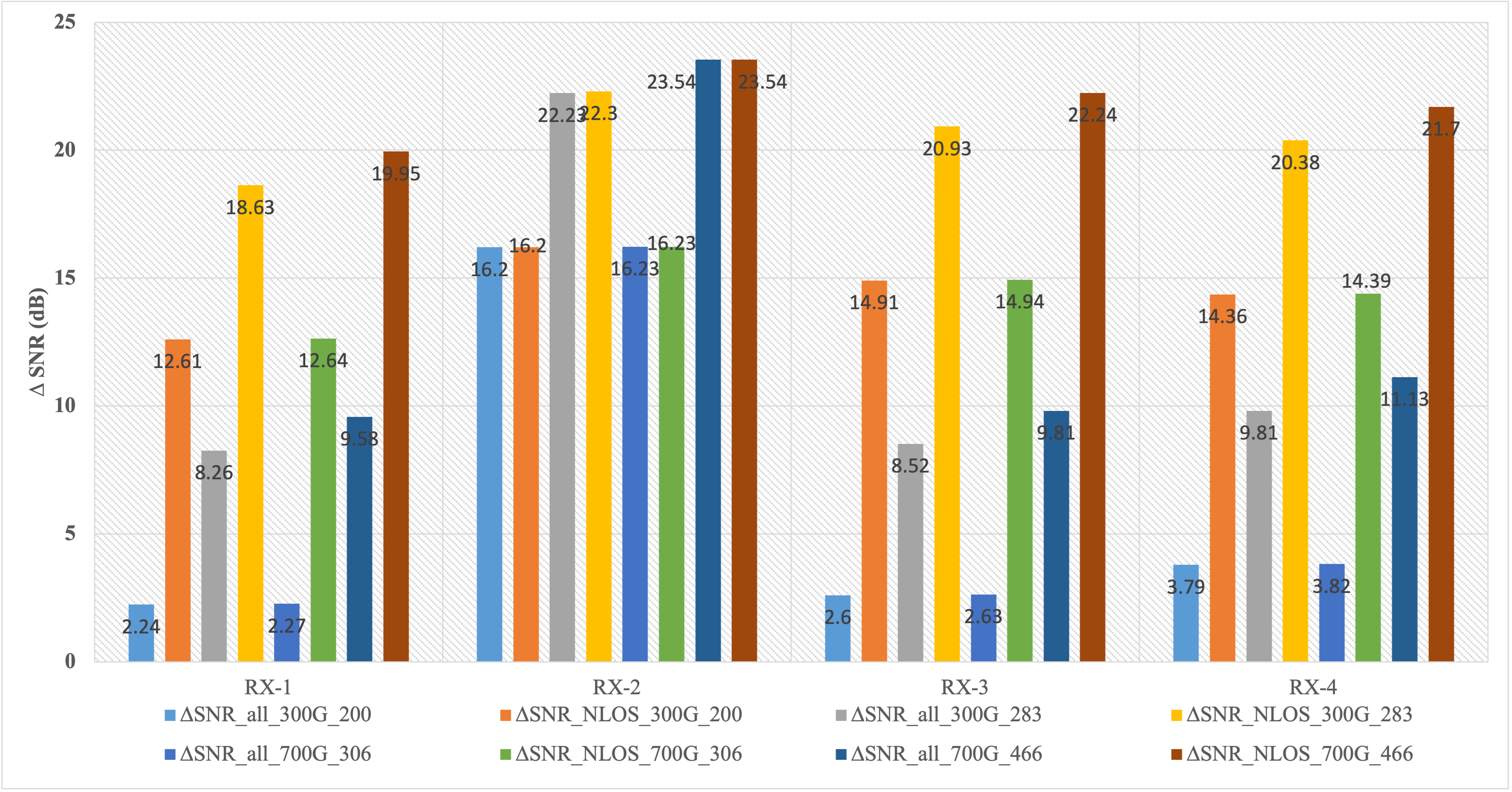}
\caption{Average SNR improvement comparison of various frequency and RIS dimension configurations, e.g. (300 GHz, $200\times200$), (300 GHz, $283\times283$), (700 GHz, $306\times306$), (700 GHz, $466\times466$), for all examined receivers.} \label{fig:SNRs}
\end{figure*}

\section{Numerical Results and Quantitative Analysis}
In this section, we apply Algorithm \ref{alg:algorithm_DRF} to the specific indoor environment in Fig.~\ref{fig:Indoor3D} with major parameters and settings illustrated in Table~\ref{Table:Simulation}. Unless specified, RIS $M$ and $N$ are both set to 200 for the analysis in the simulation. It is worth noting that both TX and RX adopt directional antennas or antenna arrays, which is a practical assumption in the THz communication. For the RIS, we set the dimension of each unit cell equal to $d_{x}$ and $d_{y}$, and make $M$ and $N$ adjustable to vary their values to unveil more relationships between them and the overall system performance. In general, a suitable RIS dimension is needed to re-establish a reflection link with smaller or equivalent path loss against the natural LOS path. At the same time, its dimension and hardware complexity should also be considered, although there is no regulation on it.       

\subsection{Quantitative Analysis of Application Scenarios at 300 GHz}
Take the case of the human blockage position (HBP) numbering ``010'' for example, as illustrated in Fig.~\ref{fig:TV} (a), through the ray-tracing simulation, two NLOS communication propagation paths are available for the TX to steer the antenna(s) maximum radiation direction. Moreover, the path loss at the two NLOS paths is 107.43 dB and 108.73 dB, respectively. Following the step of obtaining the NLOS parameters, DRF Algorithm 1 examines the available RIS(s) as shown in Fig.~\ref{fig:TV} (b) and (c), and further computes the path loss as 89.04 dB and 102.19 dB, respectively. Subsequently, DRF Algorithm 1 determines the RIS-3 as the best candidate for reconstructing a reflection link that even outperforms the path loss of LOS propagation between the TX and RX-3 (TV STB), $PL_\text{LOS}$ (91.22 dB) by 2.18 dB. Compared to the best available NLOS link without activating RISs, using RIS-3 and RIS-4 can improve the SNR by 18.39 dB and 5.24 dB, respectively.    

Furthermore, a more complete statistical data visualization of ray-tracing simulation results based on all 727 human blockage positions and RX-3 operating at 300 GHz, is illustrated in Fig.~\ref{fig:HBP_RX-3}. There are 26 HBPs where LOS communication is disrupted, and only NLOS communication is available when the RIS is not activated. On average, LOS communication occurrence is 96.42\% of total 727 HBPs with a path loss mean value, $\mu_\text{PL}$, of 91.22 dB and a standard deviation, $\sigma_\text{PL}$, of 0 dB while NLOS communication accounts for the remaining 3.58\% HBPs with (103.95 dB, 3.76 dB) for ($\mu_\text{PL}$, $\sigma_\text{PL}$). With RISs activated, there are several typical combinations as candidate solutions. The first one only relies on the RISs, and RIS-3 enables 98.35\% of the communication while the remaining 1.65\% is from RIS-4. It is worth noting that 98.35\% is the maximum ratio that RIS-3 can achieve since it is also blocked by the human blocker in the remaining 1.65\% HBPs. ($\mu_\text{PL}$, $\sigma_\text{PL}$) of RIS-3 and RIS-4 is (89.04 dB, 0) and (102.19 dB, 0), respectively. In option 2, 96.42\% of the communication still depends on the available LOS communication, while the remaining 3.58\% HBPs rely on the RIS-3, which results in (89.04 dB, 0). Finally, in option 3, RIS-3 enables 98.35\% of total HBPs while the remaining 1.65\% is enabled by the original LOS communication, which is not blocked by the human blocker. Noticeably, for the RX-3 scenario, the remaining 1.65\% HBPs that cannot be enabled by RIS, are all available for the LOS communication through ray-tracing simulations, but this may not always be valid for other receivers.

Through further computation and comparison, for RX-3, option-3 strategy stands out with 2.6 dB average overall SNR increment,~$\overline{\Delta\text{SNR}_\text{all}}$ and 14.91 dB average NLOS SNR increment, $\overline{\Delta\text{SNR}_\text{NLOS}}$, respectively. In particular, $\overline{\Delta\text{SNR}_\text{all}} = \overline{PL_\text{W/O-RIS}} - \overline{PL_\text{Opt.}}$, $\overline{\Delta\text{SNR}_\text{NLOS}} =  \overline{PL_\text{W/O-RIS}^\text{NLOS}} -\overline{PL_\text{Opt.}^\text{RIS}}$, where $ \overline{PL_\text{W/O-RIS}}$ represents the weighted average path loss of all HBPs under the selected receiver without RISs activated, $\overline{PL_\text{Opt.}}$ is the weighted average path loss of the selected option, $\overline{PL_\text{W/O-RIS}^\text{NLOS}}$ stands for the weighted average path loss of the minimum NLOS (NLOS-min) propagation occurrence without RISs, and $\overline{PL_\text{Opt.}^\text{RIS}}$ represents the the weighted average path loss of the selected RIS over the former total occurrences of NLOS-min propagation. In addition, statistical data visualization based on 300 GHz and RX-2 is given in Fig.~\ref{fig:HBP_RX-2} where LOS communication is completely unavailable.

Furthermore, all receivers, RX 1-4, have been examined at all HBPs with the results and summary presented in Table~\ref{tab:APP300G} where the best strategies resulting in the best SNR improvements are individually given. 
There are several conclusions to extract from the observation:

\begin{enumerate}
\item The proportion of LOS and NLOS occurrences varies significantly. It depends on the receiver locations, but usually, the average path loss of LOS is more than 10 dB better than the NLOS-min, which indicates that the building material absorbs a substantial portion of the EM wave energy after the reflection. 

\item With RISs activated, 761 out of 762 former NLOS scenarios can be re-established with RISs enabled artificial reflection links. The overall SNR improvement success rate is 99.87\% and RX1-4 occupies a proportion of 1.05\%, 95.4\%, 3.42\% and 0.13\%, respectively. The $\overline{\Delta\text{SNR}_\text{NLOS}}$ is at least higher than 12.6 dB for any receiver.  

\item More importantly, deploying RISs in a distributed manner is critical and can improve the overall success rate since one RIS can be blocked or even blocked with the LOS communication simultaneously in some very rare cases.

\item The SNR improvements can be scaled up with the RIS hard area. The above simulation has used a uniformed hardware area of $0.07 \times 0.07~\text{m}^2$ for each RIS when setting both $N$ and $M$ to 200, which is only half the area of a typical smartphone. More SNR improvement is expected when adopting a larger hardware area of the RIS.

\end{enumerate}

\begin{figure}
\begin{minipage}{.5\linewidth}
\centering
\subfigure[]{\includegraphics[scale=.478]{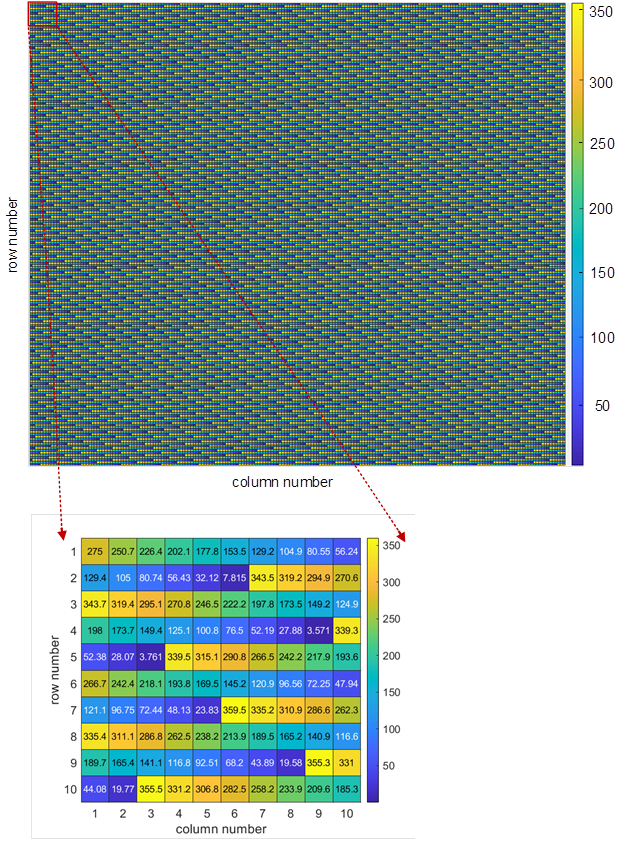}}
\end{minipage} \\
\begin{minipage}{.5\linewidth}
\centering
\subfigure[]{\includegraphics[scale=.478]{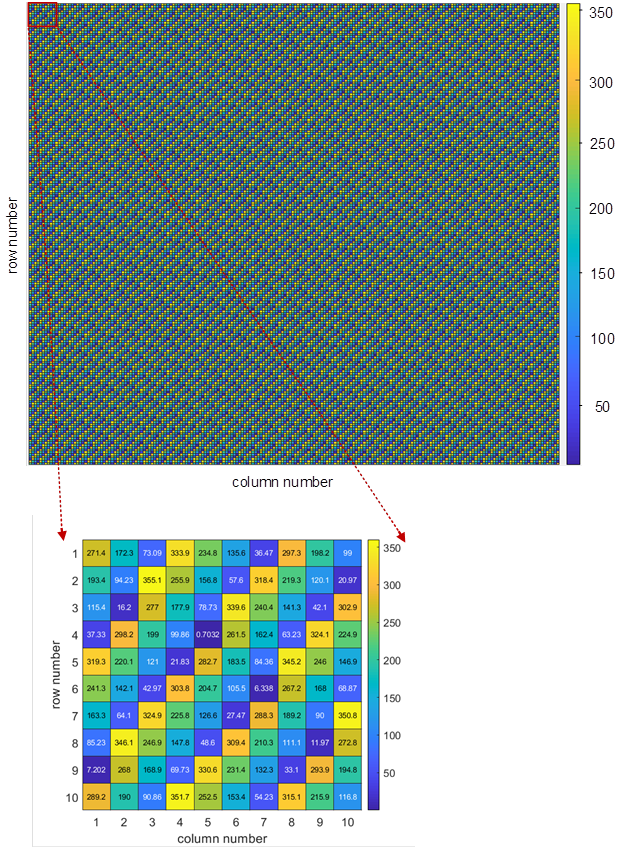}}
\end{minipage}\par\medskip
\caption{Phase shift settings of TX to RX4 link established by (a) RIS3 and (b) RIS4, both with a zoom-in illustration of the first $10 \times 10$ sub-array of RIS unit cells.}\label{fig:PS}
\end{figure}

\subsection{SNR Improvement Scaling of RIS Dimension and Frequency}

More simulations of various combinations have been conducted and analyzed at 300 GHz, and 700 GHz with different RIS dimensions. As illustrated in Fig.~\ref{fig:SNRs}, the benchmark is set to the combination of (300 GHz, $200\times200$). When the RIS hardware area is increased to double as the benchmark, i.e., $283\times283$ for $M\times N$, around 6 dB more SNR improvement gain is observed at the receiver end, which is in line with (\ref{eq:PL_Gen}), (\ref{eq:PL_FarBeam}), and (\ref{eq:PL_NearBeam}) where the path loss decreases by around 6 dB when the product of $M\times N$ is doubled. 

Furthermore, when the frequency increases from 300 GHz to 700 GHz, we need to increase $M$ and $N$ to 306 to maintain the same SNR improvement as the 300 GHz case, which is also in alignment with (\ref{eq:PL_Gen}), (\ref{eq:PL_FarBeam}), (\ref{eq:PL_NearBeam}) and propagation loss equations. The surface area of 700 GHz RIS becomes 0.046 × 0.046 m$^{2}$ which is 3/7 of the surface area of the 300 GHz RIS. If we want to maintain the same RIS-enabled path loss at 700 GHz as the 300 GHz benchmark, the surface area of the 700 GHz RIS needs to be the same as the 300 GHz case. Then, both $M$ and $N$ need to be increased to 466, which results in a further 7.3 dB increment of the SNR improvement. More data on the SNR improvement under different parameters is given in Fig.~\ref{fig:SNRs}.   

In addition, the phase shift of each unit cell in the RIS can be calculated and pre-stored for each simulated scenario. As illustrated in Fig.~\ref{fig:PS}, all phase shift values (between 0 and 360 degrees) of RIS3 and RIS4 are given, when establishing the communication links between the transmitter and the receiver RX4. It is observed that the phase shift settings for fixed receivers are pre-determined, while the mobile receivers require real-time localization and phase shift calculation. Therefore, for IoT application scenarios where the locations of IoT devices and sensors are usually not changed frequently (in seconds or minutes), phase shift settings are less complex. However, it is particularly noteworthy that, due to the high cost and challenging complexity of implementing the continuous control of the phase shifts for other application scenarios, it is cost-effective to implement discrete and finite reflection for practical IRSs \cite{Wu:PS1}. In fact, the research work \cite{Wu:PS2} has demonstrated that using 3-bit phase shifters (controlled by three positive-intrinsic-negative (PIN) diodes) is practically sufficient to achieve close-to-optimal performance with only approximately 0.2 dB power loss when the number of IRS elements is large.

The above observations can help us to determine the RIS dimension at different frequencies in different environments. The path losses of the re-constructed reflection links are mostly RIS dimension dependent, and more SNR improvements can be expected at higher frequencies when the dimension is unchanged. On the other hand, implementation should consider the hardware complexity as more RIS unit cells need to be controlled and synchronized for the higher frequency with unchanged RIS dimension. 

\subsection{Implementation and Deployment of Distributed RISs Framework}

A further discussion on the implementation and deployment feasibility of the proposed DRF is presented in this subsection. Algorithm~\ref{alg:algorithm_DRF} provides the fundamental principle of extracting the channel parameters with and without RISs' assistance and eventually determining which RIS(s) to use. Such an algorithm is practical and can be implemented on an emergent system such as a small cell or femtocell that is capable of doing integrated sensing and communications (ISAC) \cite{Tan:6G}. The 3GPP has begun a feasibility study of ISAC in release 19 \cite{2022:3GPP}, wherein various promising use cases have been discussed. The integration of dual functionality of sensing and communication is supposed to be a critical feature of the 6G radio access network (RAN) \cite{Liu:ISAC}. Although 5G NR waveform can be used to enable target localization \cite{Kanhere:ISAC}, its accuracy might be physically limited by the available bandwidth at mmWave. At THz frequencies, where a much larger bandwidth is available than its mmWave counterpart, the ISAC within the perceptive 6G networks can be enhanced and support more new services, including high-accuracy localization and tracking, simultaneous imaging and mapping, augmented human sense, and gesture and activity recognition \cite{Tan:6G}.   

\begin{figure}
\centering
\includegraphics[scale = 0.5]{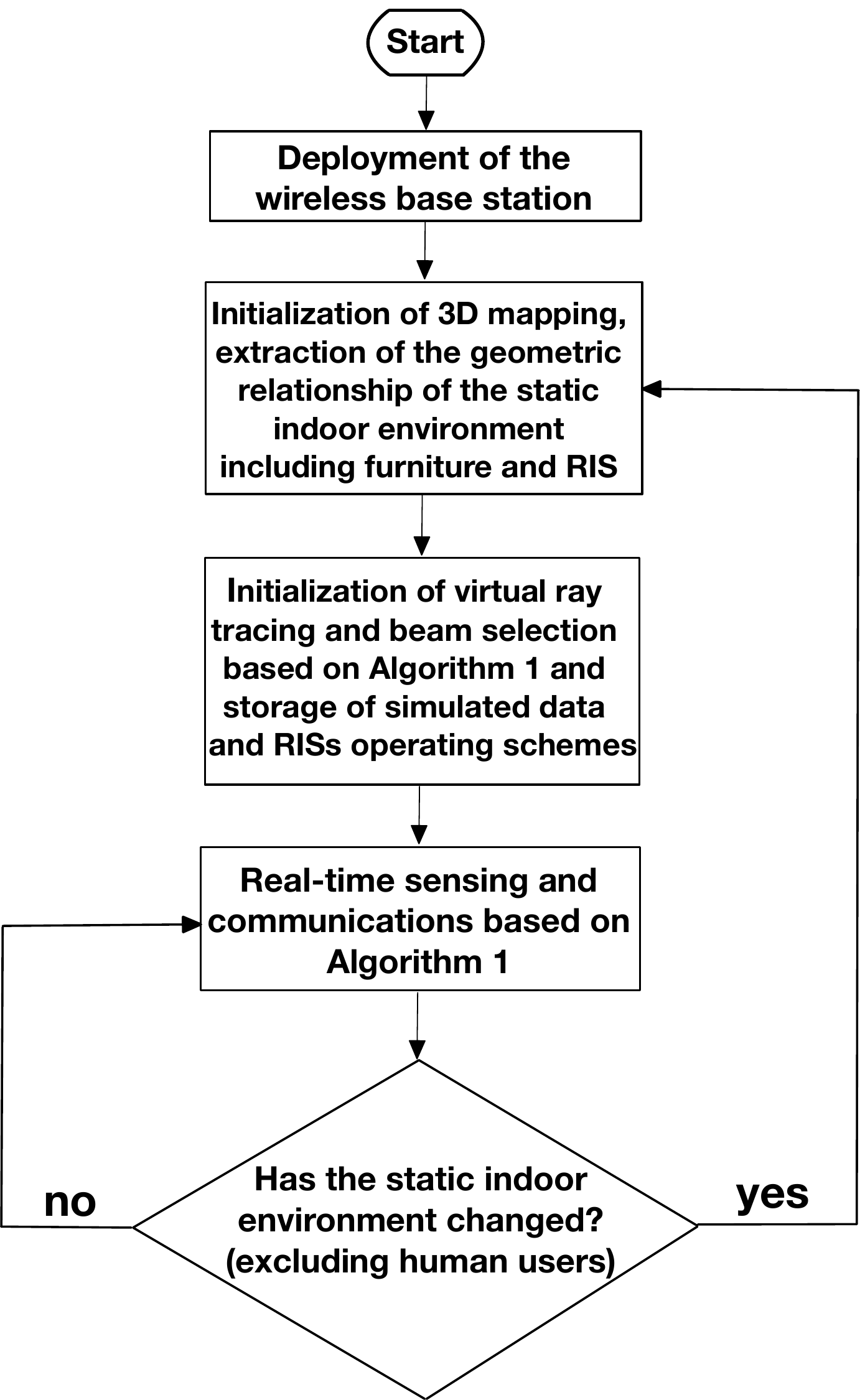}
\caption{Simplified flowchart of DRF deployment and operation.}\label{fig: FlowChart_DRF}
\end{figure}

The simplified flowchart of the DRF deployment and operation is shown in Fig.~\ref{fig: FlowChart_DRF}. For an exemplary case of typical DRF deployment, the femtocell (or the wireless access point) is initialized to obtain the 3D mapping information of the indoor environment through one or multiple high-precision/resolution sensing methods, such as RF sensing (THz), or non-RF-based sensing such as near-infrared (NIR), and 3D LiDAR sensing \cite{Yin:Lidar}, \cite{Ogunniyi:Lidar}. It is worth mentioning that the spectral analysis at THz bands (2.5 to 7.5 THz) can be utilized to understand the composition of materials \cite{Tan:6G}, which further facilitates more intelligent and efficient indoor sensing and communication. In this step, all needed static indoor geometric information is extracted. Subsequently, the femtocell can operate the massive training based on Algorithm~\ref{alg:algorithm_DRF}, either on-device or through a cloud. Eventually, the femtocell can learn all possible combinations of human blocker(s) positions plus receivers' coordinates and know what strategy should be applied to the situation whenever and wherever human blocker(s) are. 

Furthermore, each strategy should also include the phase shift settings for the RISs. All these settings are stored in something similar to a look-up table and ready for inference after the training. Then in a realistic DRF operation, the real-time coordinates of both receiver(s) and human blockers are timely and precisely obtained through sensing, localization, and computer vision techniques. Those coordinates immediately trigger the femtocell to beamform the signal to the best direction with the best strategy that is already pre-known. As a result, the overhead, latency, and power consumption can be largely minimized while the SNR is improved, as compared to the existing beamforming protocols \cite{Hosoya:60G}.    

It is also envisioned that, when the presumably static environment changes, e.g. furniture and objects are rearranged, the femtocell needs to redo the 3D mapping, virtual ray tracing, etc. training to upgrade the overall strategy. The new inference models can be configured with an automatic update based on an hourly, daily, or even weekly pace, which depends on the realistic user experience, quality of service, and technical feasibility. Last but not least, it is noteworthy that a RIS itself may consume significantly lower power than an amplifier and forward (AF) relay with power-hungry active RF components inside it. Moreover, a RIS system can be considered a quasi-passive, virtually extended lens antenna system since it needs to be connected to the femtocell through either wireless or wired. This means a RIS controller accommodating the control signal interface and phase-shifters, etc., requires a much lower power. 

On the one hand, the power consumption of a RIS system is related to the frequency of the phase-shift setting, which requires the activation of the control link and tuning RISs’ phase shifts. When one specific phase shift matrix is applied for one targeted receiver that is assumed to be temporarily static, the theoretical DC power consumption is almost 0 unless the femtocell needs to frequently change its phases to reflect the wireless signal to another receiver or a moving receiver.

On the other hand, a suitable machine learning (ML) framework can be developed and deployed to improve overall energy efficiency and system performance. For example, the deep neural network can be combined with sensing/localization techniques to improve positioning accuracy \cite{Lu:Deep}. Moreover, a deep-reinforcement-learning (DRL)-based unsupervised wireless-localization method can waive the acquisition of the time-consuming and costly location labels \cite{Li:Deep}. Also, a deep neural network trained offline using the unsupervised learning mechanism can perform real-time phase shift configuration for the RIS while maintaining a decent rate of performance \cite{Gao:Unsupervised}. Furthermore, in a practical 6G ISAC system where sensing and localizing a mobile user in a real-time manner can be energy/resource consuming, predicting human behavior, motion, and gait through a suitable ML framework \cite{Yi:Behavior}, \cite{Xing:Comprehensive} can enable less frequent sensing and improve the overall energy efficiency of the proposed DRF. Further, a fusion of federated learning and edge computing with the DRF can help minimize the data privacy issue and enhance user privacy protection \cite{Lim:Federated}.    





\section{Conclusion}
Based on the review of the most recent research progress on the reconfigurable intelligent surface-assisted wireless communications, we have proposed a distributed RISs framework to enable a better quality of service in THz indoor communications. In particular, it can facilitate energy-efficient IoT applications. By conducting massive ray-tracing simulations of a realistic 3D indoor environment with mobile human blockers and various receivers and analyzing a large amount of obtained data, the proposed DRF algorithm can fast learn and thoroughly evaluate the entire dynamic environment. Eventually, it can determine the best strategy for THz data transmission. The four distributed RISs-based DRF's overall success rate is 99.87\% for re-establishing reflection links of former NLOS communication without RISs scenarios where indoor building materials can significantly absorb and attenuate the THz signals more than 10 dB. Consequently, the overall system energy efficiency is improved by substituting the natural reflections (by objects or humans) with the RIS's artificial reflections that almost have no DC power consumption for maintaining the phase shifts with static bias voltages.    

Furthermore, the SNR improvements for both 300 GHz and 700 GHz, under different RIS hardware dimensions, are also quantitatively analyzed. The SNR improvement scaling law is unveiled, and practical RIS hardware system design constraints are discussed. In general, the RIS hardware area used for this research is practically small and feasible for engineering purposes. Eventually, we further investigate and present the main procedures and critical details of deploying and operating the DRF-based THz communication system to balance energy efficiency, cost-effectiveness, and quality of service.  

The future work lies in several aspects to enable better energy efficiency, which includes integrating RISs into more vertical application scenarios, developing machine learning frameworks for the RIS system, investigating the deployment of the DRF for various indoor environments, and extending DRF for more challenging environments such as outdoor and high-dense urban scenarios.

\end{document}